\documentclass[a4paper,11pt]{article}
\pdfoutput=1
\usepackage{jheppub}

\usepackage{amssymb,amsmath,amsfonts}
\usepackage[normalem]{ulem}
\usepackage[utf8x]{inputenc}
\usepackage{slashed}
\usepackage{graphicx}
\usepackage{tabularx}
\usepackage{here}
\usepackage{color}
\usepackage{csquotes} 
\usepackage{comment}
\usepackage{mathrsfs}
\usepackage{float}
\usepackage{ascmac}
\usepackage{multirow}
\usepackage{longtable}
\usepackage{bm}
\usepackage{ulem}

\usepackage[italicdiff]{physics}

\makeatletter
\newcommand*\rel@kern[1]{\kern#1\dimexpr\macc@kerna}
\newcommand*\widebar[1]{%
  \begingroup
  \def\mathaccent##1##2{%
    \rel@kern{0.8}%
    \overline{\rel@kern{-0.8}\macc@nucleus\rel@kern{0.2}}%
    \rel@kern{-0.2}%
  }%
  \macc@depth\@ne
  \let\math@bgroup\@empty \let\math@egroup\macc@set@skewchar
  \mathsurround\z@ \frozen@everymath{\mathgroup\macc@group\relax}%
  \macc@set@skewchar\relax
  \let\mathaccentV\macc@nested@a
  \macc@nested@a\relax111{#1}%
  \endgroup
}
\makeatother

\numberwithin{equation}{section}

\preprint{
\begin{minipage}{5cm}
\small
\flushright
EPHOU-25-007\\
KYUSHU-HET-321
\end{minipage}}

\title{Lepton mass textures from non-invertible multiplication rules}

\author{Tatsuo Kobayashi$^{1}$,} 
\author{Hajime Otsuka$^{2}$,} 
\author{Morimitsu Tanimoto$^{3}$, and} 
\author{Haruki Uchida$^{1}$}
\affiliation{
$^1$Department of Physics, Hokkaido University, Sapporo 060-0810, Japan}
\affiliation{
$^2$Department of Physics, Kyushu University, 744 Motooka, Nishi-ku, Fukuoka 819-0395, Japan}
\affiliation{
$^3$Department of Physics, Niigata University, Ikarashi 2-8050, Niigata 950-2181, Japan}
\emailAdd{kobayashi@particle.sci.hokudai.ac.jp}
\emailAdd{otsuka.hajime@phys.kyushu-u.ac.jp}
\emailAdd{morimitsutanimoto@yahoo.co.jp}
\emailAdd{haru-uchida@particle.sci.hokudai.ac.jp}

\abstract{
We study the lepton mass textures, which are derived by $\mathbb{Z}_2$ gauging of $\mathbb{Z}_M$ symmetries.
We can obtain various textures for the Yukawa couplings in the charged lepton sector, but the patterns of neutrino mass matrices are limited.
All the obtained textures can not be realized by group-theoretical symmetries, and certain textures can lead to realistic results.
}

\makeatletter
\gdef\@fpheader{}
\makeatother

\begin{document}

\maketitle

\section{Introduction}
\label{sec:Intro}

Symmetries are quite important in physics.
In particle physics, symmetries determine which couplings are allowed or forbidden.
That is the coupling selection rules.
We often use coupling selection rules due to group theory.
Recently, the concept of symmetries was generalized and symmetries without a group structure were also studied.
(See for reviews on non-invertible symmetries Refs.~\cite{Gomes:2023ahz,Schafer-Nameki:2023jdn,Bhardwaj:2023kri,Shao:2023gho}.)

One of the important mysteries in particle physics is the origin of the flavor structure, i.e., mass hierarchies, mixing angles, and CP phases of quarks and leptons.
One type of approach to address this mystery is to impose group-theoretical flavor symmetries including Abelian and non-Abelian, and continuous and discrete groups \cite{Froggatt:1978nt,Altarelli:2010gt,Ishimori:2010au,Hernandez:2012ra,King:2013eh,King:2014nza,Petcov:2017ggy,Kobayashi:2022moq}.
Experimental values of fermion masses, mixing angles, and CP phases do not have exact group-theoretical symmetries.
Group-theoretical flavor symmetries must be broken to derive realistic values by vacuum expectation values of scalar fields, i.e., flavons.
In this sense, how to break flavor symmetries is important.
Recently, modular flavor symmetries were also studied intensively \cite{Feruglio:2017spp}.
In those models, Yukawa couplings also transform non-trivially under flavor symmetries such as $S_3, A_4, S_4, A_5$ \cite{Kobayashi:2018vbk,Feruglio:2017spp,Penedo:2018nmg,Novichkov:2018nkm}, which is different from conventional flavor symmetric models.
(See for reviews of modular flavor symmetric models Refs.~\cite{Kobayashi:2023zzc,Ding:2023htn}.)

Another approach to address the flavor mystery is to assume specific textures of quark and lepton mass matrices.
This approach was started by Weinberg \cite{Weinberg:1977hb}.
Many interesting textures have been proposed \cite{Fritzsch:1977vd,Fritzsch:1979zq,Ramond:1993kv,Fritzsch:2002ga,Xing:2015sva,Frampton:2002yf,Kageyama:2002zw}.
The texture zero approach is still promising to address the flavor mystery.
However, it has no definite theoretical reason why we can start with specific textures.

As mentioned above, symmetries were generalized.
Over the past years, non-invertible symmetries have been applied in several topics of particle physics \cite{Choi:2022jqy,Cordova:2022fhg,Cordova:2022ieu,Cordova:2024ypu,Delgado:2024pcv,Suzuki:2025oov,Liang:2025dkm}.
In particular, non-invertible selection rules derived from magnetized compactification \cite{Kobayashi:2024yqq,Funakoshi:2024uvy} were applied to understand the origin of the flavor structure very recently \cite{Kobayashi:2024cvp,Kobayashi:2025znw}.
That is $\mathbb{Z}_2$ gauging of $\mathbb{Z}_M$ symmetries.
By use of $\mathbb{Z}_2$ gauging, one can derive interesting Yukawa textures, which have been proposed without definite theoretical reasons.
For example, the nearest neighbor interaction texture \cite{Branco:1988iq} can be obtained \cite{Kobayashi:2024cvp}.
A new type of Yukawa textures has been derived and it can lead to realistic values of quark masses and mixing angles as well as the CP phase \cite{Kobayashi:2025znw}.

Our purpose of this paper is to extend the analyses in Refs.~\cite{Kobayashi:2024cvp,Kobayashi:2025znw} to the lepton sector.
We study the neutrino mass matrix derived from the Weinberg operator in addition to the Yukawa matrix of the charged lepton sector. 
In Refs.~\cite{Kobayashi:2024cvp,Kobayashi:2025znw}, Yukawa textures in the quark sector were analyzed. 
The analysis on the Yukawa textures in the charged lepton sector is similar to those in Refs.~\cite{Kobayashi:2024cvp,Kobayashi:2025znw}. 
However, the study on the neutrino mass matrices are a new analysis here, and we study possible combinations between neutrino mass textures and Yukawa textures in the charged lepton sector.

This paper is organized as follows.
In section \ref{sec:Generic}, we briefly review the scenario that non-trivial multiplication rules lead to a new type of coupling selection rules.
In section \ref{sec:Z2-gauging}, we give a review on $\mathbb{Z}_2$ gauging of $\mathbb{Z}_M$ symmetries.
In section \ref{sec:lepton}, we study which textures in the lepton sector are derived by $\mathbb{Z}_2$ gauging of $\mathbb{Z}_M$ symmetries with $M=3,4,5$.
Phenomenological aspects of obtained textures are discussed in section \ref{sec:pheno}. 
Section \ref{sec:con} is our conclusion.

\section{Coupling selections due to non-trivial multiplication rules}
\label{sec:Generic}

Here, we review on coupling selection rules  from non-trivial multiplication rules comparing with coupling selection rules due to group theory.

In a group theory, elements $a,b,c$ in the group $G$ satisfies the multiplication law,
\begin{align}
    ab=c.
\end{align}
Note that the element appearing in the right hand side is unique.
In $G$-invariant field theory, each field is assigned to a (representation of) group element as $\phi_a, \phi_b, \phi_c$.
The process $\phi_a + \phi_b \to \phi_c$ is allowed when $ab=c$.
That is the charge conservation law when $G$ is Abelian.
On the other hand, the process $\phi_a + \phi_b \to \phi_d$ is forbidden when $ab \neq d$.
A different field $\phi'_c$ can correspond to the same 
element $c$ as $\phi_c$, and the process $\phi_a + \phi_b \to \phi'_c$ is allowed by group theory.

We consider the set of elements $\{ U_i\}$, which may be operators.
Underlying theory may lead to non-trivial multiplication rules,
\begin{align}
    U_i U_j = \sum_{k}c_{ijk}U_k.
\end{align}
Here, the elements appearing in the right hand side are not unique, and these multiplication rules are non-invertible.
Each mode $\phi_i$ corresponds to an element $U_i$.
Coupling selection rules of $\phi_i$ in low energy effective field theory is determined by the above multiplication laws.
The process $\phi_i + \phi_j \to \phi_k$ can occur 
when $c_{ijk} \neq 0$.
A different mode $\phi'_k$ can correspond to the same $U_k$, and the process $\phi_i + \phi_j \to \phi'_k$ can occur when  $c_{ijk} \neq 0$.

\section{Coupling selection rules due to $\mathbb{Z}_2$ gauging of $\mathbb{Z}_M$ symmetries}
\label{sec:Z2-gauging}

Here, we review about $\mathbb{Z}_2$ gauging of $\mathbb{Z}_M$ symmetries.

We start with an illustrating model for $\mathbb{Z}_2$ gauging of $\mathbb{Z}_M$ symmetries \cite{Kobayashi:2024yqq,Funakoshi:2024uvy}. 
Certain compactifications such as magnetized compactifications of higher-dimensional Yang-Mills theory lead to the $\mathbb{Z}_M$ symmetry \cite{Abe:2009vi,Berasaluce-Gonzalez:2012abm,Marchesano:2013ega}. 
Then, the massless modes $\varphi_k$ transform as 
\begin{align}
    \varphi_k \to g^k\varphi_k,
\end{align}
under the $\mathbb{Z}_M$ symmetry, where $g=e^{2\pi i/M}$.
Here, we consider geometrical $\mathbb{Z}_2$ orbifolding of the above compactification.
The $\mathbb{Z}_2$ invariant modes are written by \cite{Abe:2008fi}
\begin{align}
    \phi_k=\varphi_k+\varphi_{M-k},
\end{align}
up to normalization.
The $\mathbb{Z}_M$ symmetry is broken.
However, certain coupling selection rules, which originate from the $\mathbb{Z}_M$ charge conservation, remain in low energy effective field theory.
The mode $\phi_k$ behaves as if it has both $k$ and $(M-k)$ charges under $\mathbb{Z}_M$ at the same time.

Here, we explain $\mathbb{Z}_2$ gauging of $\mathbb{Z}_M$ symmetries based on the above illustrating model.
The multiplication law of the $\mathbb{Z}_M$ symmetry is written by 
\begin{align}
    g^{k_1}g^{k_2}=g^{k_1+k_2}.
\end{align}
The $n$-point couplings of $\varphi_{k_1},\cdots , \varphi_{k_n}$ are allowed if 
\begin{align}
    g^{k_1}\cdots g^{k_n}=g^0,
\end{align}
where $\varphi_{i}$ corresponds to the element $g^{k_i}$ $(i=1,\cdots,n)$.

Here, we consider the following automorphisms:
\begin{align}
    eg^{k}e^{-1},\qquad rg^{k}r^{-1}=g^{M-k},
\end{align}
where the latter corresponds to the outer automorphism of $\mathbb{Z}_M$, while the former is trivial one.
By using them, we define the following class:
\begin{align}
    [g^k]=\{hg^kh^{-1} ~|~ h=e, r \}.
\end{align}
That is identification of two $\mathbb{Z}_M$ charges. This procedure is called as $\mathbb{Z}_2$ gauging of $\mathbb{Z}_M$ symmetry, and it corresponds to geometrical $\mathbb{Z}_2$ orbifolding. 
Indeed, the $\mathbb{Z}_2$ even modes on $T^2/\mathbb{Z}_2$ orbifold are labeled by this class~\cite{Kobayashi:2024yqq}. 
We consider the case with $M > 2$ because the $\mathbb{Z}_2$ symmetry remains when $M=2$.
These classes satisfy the following multiplication rules:
\begin{align}
\label{eq:multi-law}
    [g^{k_1}][g^{k_2}]=[g^{k_1+k_2}]+[g^{{k_1}-{k_2}}].
\end{align}

Each mode $\phi_k$ corresponds to a class $[g^k]$.
The two point couplings (mass terms) of $\phi_{k_1}$ and $\phi_{k_2}$ are allowed if the identity class $[g^0]$ appears in the right hand side of Eq.~(\ref{eq:multi-law}).
The condition on allowed two point couplings is written by 
\begin{align}
\label{eq:2-point-rule}
    \pm k_1 \pm k_2 = 0 ~~~({\rm mod~~}M).
\end{align}
That means that the mass terms and kinetic terms are allowed between the same class $[k_1]=[k_2]$.

Similarly, we can write the condition on allowed three point couplings among $\phi_1, \phi_2, \phi_3$ as 
\begin{align}
    \pm k_1 \pm k_2 \pm k_3  =0 ~~~({\rm mod~~}M).
\end{align}
Moreover, the $n$ point couplings among $\phi_1,\cdots, \phi_n$ are allowed if the following condition:
\begin{align}
    \sum_{i=1}^n \pm k_i  =0 ~~~({\rm mod~~}M)
\end{align}
 is satisfied.

For example, when $M=3$ there are two classes, $[g^0]$ and $[g^1]$.
They satisfy the following fusion rules:
\begin{align}
    [g^0][g^0]=[g^0], \qquad [g^0][g^1]=[g^1],\qquad 
    [g^1][g^1]=[g^0]+[g^1],
\end{align}
which are the so-called Fibonacci fusion rules.
When $M=4$, there are three classes, $[g^0]$, $[g^1]$, and $[g^2]$.
They satisfy the following fusion rules:
\begin{align}
    &[g^0][g^0]=[g^0], \qquad [g^0][g^1]=[g^1],\qquad 
    [g^0][g^2]=[g^2], \notag \\
    &[g^1][g^2]=[g^1], \qquad [g^2][g^2]=[g^0], \qquad 
    [g^1][g^1]=[g^0]+[g^2],
\end{align}
which correspond to the Ising fusion rules.

\section{Lepton mass matrices}
\label{sec:lepton}

Here, we study which patterns of lepton mass matrices can be derived by $\mathbb{Z}_2$ gauging of $\mathbb{Z}_M$ with $M=3,4,5$.

\subsection{Neutrino mass matrix by Weinberg operators}
\label{sec:Weinberg0op}

Let us focus on neutrino mass matrix, which is derived by the Weinberg operators,
\begin{align}
    \frac{{C_{\mathrm{W}}}^{ij}}{\Lambda}(L_i\varepsilon H)C(L_j \varepsilon H),
\end{align}
including three generations of the left-handed leptons $L_i$ and the Higgs field $H$, where $C$ denotes the charge conjugate, and $\Lambda$ denotes a cut-off scale.
Note that the field $\phi_k$ and its conjugate correspond to the same class $[g^k]$.
In addition, the multiplication of $[g^k][g^k]$ corresponding to $L_i L_i$ always includes $[g^0]$ in Eq.~(\ref{eq:multi-law}) as the two-point couplings Eq.~\eqref{eq:2-point-rule}.
Similarly, the multiplication of classes corresponding to $HH$ always includes $[g^0]$.
That implies that the diagonal elements of ${C_{\mathrm{W}}}^{ij}$ are always allowed by our selection rules.
In Ref.~\cite{Frampton:2002yf}, the neutrino mass textures, $A_{1,2}$, $B_{1,2,3,4}$, $C$, were studied.
They include texture zeros in the diagonal entries in the basis that the charged lepton mass matrix is diagonal.
Our selection rules can not derive those textures.
In what follows, we study explicitly the textures of ${C_{\mathrm{W}}}^{ij}$ by $\mathbb{Z}_2$ gauging of $\mathbb{Z}_M$ with $M=3,4,5$.

\subsubsection{$M=3$}
\label{sec:M=3}

When $M=3$, there are two classes, $[g^0]$ and $[g^1]$.
If all three generations of $L_i$ correspond to the same class, every entry of ${C_{\mathrm{W}}}^{ij}$ is allowed.
That is rather trivial.
We study other assignments of classes to $L_i$.
The three generations of left-handed leptons $L_i$
are assigned to classes in the following two patterns:
\begin{align}
\label{eq:flavor-assign-M=3}
    \text{i} : ([g^0],[g^1],[g^1]), \qquad 
    \text{ii} : ([g^0],[g^0],[g^1]),
\end{align}
including their permutations.
The Higgs mode can correspond to either $[g^0]$ or $[g^1]$.
Then, we can examine which entries of ${C_{\mathrm{W}}}^{ij}$ are allowed by our selection rules.
Results are shown in Table~\ref{tab:C-M=3}.
These textures can be understood effectively by the $\mathbb{Z}_2$ symmetry.
For example, when the Higgs field has the class $[g^0]$, 
the textures can be obtained by the assignment that 
the modes with $[g^0]$ and $[g^1]$ correspond to $\mathbb{Z}_2$ even and odd, respectively.

\begin{table}[H]
    \centering
    \begin{tabular}{|c|c|c|} \hline
      Flavor  & Higgs $[g^0]$ & Higgs $[g^1]$\\ \hline
\begin{tabular}{l}
(i)\,:\,$([g^0],[g^1],[g^1])$ 
\end{tabular} 
& $\begin{pmatrix}
* & 0 & 0 \\
0 & * & * \\
0 & * & *
\end{pmatrix}$
& $\begin{pmatrix}
* & * & * \\
* & * & * \\
* & * & *
\end{pmatrix}$  \\ \hline
\begin{tabular}{l}
(ii)\,:\,$([g^0],[g^0],[g^1])$ 
\end{tabular}   
&  $\begin{pmatrix}
* & * & 0 \\
* & * & 0 \\
0 & 0 & *
\end{pmatrix}$ & $\begin{pmatrix}
* & * & * \\
* & * & * \\
* & * & *
\end{pmatrix}$\\ \hline
    \end{tabular}
    \caption{${C_{\mathrm{W}}}^{ij}$ matrices for $M=3$. The asterisk symbols "$*$" denote non-vanishing elements.}
    \label{tab:C-M=3}
\end{table}

\subsubsection{$M=4$}
\label{sec:M=4}

We study the case with $M=4$, where 
there are three classes $[g^0],\ [g^1],\ [g^2]$.
As mentioned for $M=3$, when three generations correspond to the same class, all entries of the mass matrix are allowed.
Let us focus on other assignments.
Table~\ref{tab:Flavor-M=4} shows flavor assignments of classes.  
By use of these flavor assignments, we examine the texture structure of ${C_{\mathrm{W}}}^{ij}$. 
Results are shown in Table~\ref{tab:C-M=4}.
The obtained textures can be understood effectively by 
proper $\mathbb{Z}_n$ symmetries.

\begin{table}[H]
\scalebox{0.705}{
    \begin{tabular}{|c|c|c|c|c|c|c|c|} \hline
        Flavor & \text{i} & \text{ii} & \text{iii} & \text{iv} & \text{v} & \text{vi} & \text{vii}\\ \hline
        & $([g^0],[g^1],[g^2])$ & $([g^0],[g^0],[g^1])$ & $([g^0],[g^0],[g^2])$ & $([g^0],[g^1],[g^1])$ & $([g^0],[g^2],[g^2])$ & $([g^1],[g^1],[g^2])$ & $([g^1],[g^2],[g^2])$ \\ \hline
    \end{tabular}
    }
    \caption{Assignments for three generations of left-handed leptons $L_i$ in $M=4$.}
    \label{tab:Flavor-M=4}
\end{table}

\begin{table}[H]
    \centering
    \begin{tabular}{|c|c|c|c|} \hline
      Flavor  & Higgs $[g^0]$ & Higgs $[g^1]$ & Higgs $[g^2]$\\ \hline
\begin{tabular}{l}
(i)\,:\,$([g^0],[g^1],[g^2])$ 
\end{tabular}
& 
$\begin{pmatrix}
* & 0 & 0 \\
0 & * & 0 \\
0 & 0 & *
\end{pmatrix}$
& 
$\begin{pmatrix}
* & 0 & * \\
0 & * & 0 \\
* & 0 & *
\end{pmatrix}$
&
$\begin{pmatrix}
* & 0 & 0 \\
0 & * & 0 \\
0 & 0 & *
\end{pmatrix}$
\\ \hline

\begin{tabular}{l}
(ii)\,:\,$([g^0],[g^0],[g^1])$ 
\end{tabular}
&  
$\begin{pmatrix}
* & * & 0 \\
* & * & 0 \\
0 & 0 & *
\end{pmatrix}$ 
& 
$\begin{pmatrix}
* & * & 0 \\
* & * & 0 \\
0 & 0 & *
\end{pmatrix}$
&
$\begin{pmatrix}
* & * & 0 \\
* & * & 0 \\
0 & 0 & *
\end{pmatrix}$\\ \hline

\begin{tabular}{l}
(iii)\,:\,$([g^0],[g^0],[g^2])$ 
\end{tabular}
&
$\begin{pmatrix}
* & * & 0 \\
* & * & 0 \\
0 & 0 & *
\end{pmatrix}$ 
& 
$\begin{pmatrix}
* & * & * \\
* & * & * \\
* & * & *
\end{pmatrix}$
&
$\begin{pmatrix}
* & * & 0 \\
* & * & 0 \\
0 & 0 & *
\end{pmatrix}$\\ \hline

\begin{tabular}{l}
(iv)\,:\,$([g^0],[g^1],[g^1])$ 
\end{tabular}
&
$\begin{pmatrix}
* & 0 & 0 \\
0 & * & * \\
0 & * & *
\end{pmatrix}$ 
& 
$\begin{pmatrix}
* & 0 & 0 \\
0 & * & * \\
0 & * & *
\end{pmatrix}$
&
$\begin{pmatrix}
* & 0 & 0 \\
0 & * & * \\
0 & * & *
\end{pmatrix}$\\ \hline

\begin{tabular}{l}
(v)\,:\,$([g^0],[g^2],[g^2])$ 
\end{tabular}
&
$\begin{pmatrix}
* & 0 & 0 \\
0 & * & * \\
0 & * & *
\end{pmatrix}$ 
& 
$\begin{pmatrix}
* & * & * \\
* & * & * \\
* & * & *
\end{pmatrix}$
&
$\begin{pmatrix}
* & 0 & 0 \\
0 & * & * \\
0 & * & *
\end{pmatrix}$\\ \hline

\begin{tabular}{l}
(vi)\,:\,$([g^1],[g^1],[g^2])$ 
\end{tabular}
&
$\begin{pmatrix}
* & * & 0 \\
* & * & 0 \\
0 & 0 & *
\end{pmatrix}$ 
& 
$\begin{pmatrix}
* & * & 0 \\
* & * & 0 \\
0 & 0 & *
\end{pmatrix}$
&
$\begin{pmatrix}
* & * & 0 \\
* & * & 0 \\
0 & 0 & *
\end{pmatrix}$\\ \hline

\begin{tabular}{l}
(vii)\,:\,$([g^1],[g^2],[g^2])$ 
\end{tabular}
&
$\begin{pmatrix}
* & 0 & 0 \\
0 & * & * \\
0 & * & *
\end{pmatrix}$ 
& 
$\begin{pmatrix}
* & 0 & 0 \\
0 & * & * \\
0 & * & *
\end{pmatrix}$
&
$\begin{pmatrix}
* & 0 & 0 \\
0 & * & * \\
0 & * & *
\end{pmatrix}$\\ \hline
    \end{tabular}
    \caption{${C_{\mathrm{W}}}^{ij}$ matrices for $M=4$.  The asterisk symbols "$*$" denote non-vanishing elements.}
    \label{tab:C-M=4}
\end{table}

\subsubsection{$M=5$}
\label{sec:M=5}

We study the case with $M=5$, where 
there are three classes $[g^0],\ [g^1],\ [g^2]$.
As mentioned for $M=3$, when three generations correspond to the same class, all entries of the mass matrix are allowed.
We focus on other assignments.
Table \ref{tab:C-M=5} shows flavor assignments of classes.
By use of these flavor assignments, we examine the texture structure of ${C_{\mathrm{W}}}^{ij}$. 
Results are shown in Table~\ref{tab:C-M=5}.
Most of the obtained textures can be understood effectively by proper $\mathbb{Z}_n$ symmetries.
However, one can not derive the textures of the flavor assignment $[g^0][g^1][g^2]$ and the Higgs modes with $[g^1]$ and $[g^2]$ by conventional group-like $\mathbb{Z}_n$ symmetries. 
Suppose that $(i,i)$, $(i,j)$, and $(k,i)$ entries are allowed by $\mathbb{Z}_n$ symmetry.
That implies that $i$-th, $j$-th, and $k$-th generations have the same $\mathbb{Z}_n$ charge.
Then, the $(k,j)$ entry must be allowed by $\mathbb{Z}_n$ symmetry.
On the other hand, if the $(k,j)$ entry is forbidden, such a texture can not be realized by $\mathbb{Z}_n$ symmetry. 
This property is the same in Yukawa matrices.\footnote{These Yukawa textures and neutrino mass matrices may be realized by breaking $\mathbb{Z}_n$ symmetries in a proper way.}

\begin{table}[H]
\scalebox{0.705}{
    \begin{tabular}{|c|c|c|c|c|c|c|c|} \hline
        Flavor & \text{i} & \text{ii} & \text{iii} & \text{iv} & \text{v} & \text{vi} & \text{vii}\\ \hline
        & $([g^0],[g^1],[g^2])$ & $([g^0],[g^0],[g^1])$ & $([g^0],[g^0],[g^2])$ & $([g^0],[g^1],[g^1])$ & $([g^0],[g^2],[g^2])$ & $([g^1],[g^1],[g^2])$ & $([g^1],[g^2],[g^2])$ \\ \hline
    \end{tabular}
    }
    \caption{Assignments for three generations of left-handed leptons $L_i$ in $M=5$.}
    \label{tab:Flavor-M=5}
\end{table}

\begin{table}[H]
    \centering
    \begin{tabular}{|c|c|c|c|} \hline
      Flavor  & Higgs $[g^0]$ & Higgs $[g^1]$ & Higgs $[g^2]$\\ \hline
\begin{tabular}{l}
(i)\,:\,$([g^0],[g^1],[g^2])$ 
\end{tabular} 
& 
$\begin{pmatrix}
* & 0 & 0 \\
0 & * & 0 \\
0 & 0 & *
\end{pmatrix}$
& 
$\begin{pmatrix}
* & 0 & * \\
0 & * & * \\
* & * & *
\end{pmatrix}$
&
$\begin{pmatrix}
* & * & 0 \\
* & * & * \\
0 & * & *
\end{pmatrix}$
\\ \hline

\begin{tabular}{l}
(ii)\,:\,$([g^0],[g^0],[g^1])$ 
\end{tabular}  
&  
$\begin{pmatrix}
* & * & 0 \\
* & * & 0 \\
0 & 0 & *
\end{pmatrix}$ 
& 
$\begin{pmatrix}
* & * & 0 \\
* & * & 0 \\
0 & 0 & *
\end{pmatrix}$
&
$\begin{pmatrix}
* & * & * \\
* & * & * \\
* & * & *
\end{pmatrix}$\\ \hline

\begin{tabular}{l}
(iii)\,:\,$([g^0],[g^0],[g^2])$ 
\end{tabular}
&
$\begin{pmatrix}
* & * & 0 \\
* & * & 0 \\
0 & 0 & *
\end{pmatrix}$ 
& 
$\begin{pmatrix}
* & * & * \\
* & * & * \\
* & * & *
\end{pmatrix}$
&
$\begin{pmatrix}
* & * & 0 \\
* & * & 0 \\
0 & 0 & *
\end{pmatrix}$\\ \hline

\begin{tabular}{l}
(iv)\,:\,$([g^0],[g^1],[g^1])$ 
\end{tabular}
&
$\begin{pmatrix}
* & 0 & 0 \\
0 & * & * \\
0 & * & *
\end{pmatrix}$ 
& 
$\begin{pmatrix}
* & 0 & 0 \\
0 & * & * \\
0 & * & *
\end{pmatrix}$
&
$\begin{pmatrix}
* & * & * \\
* & * & * \\
* & * & *
\end{pmatrix}$\\ \hline

\begin{tabular}{l}
(v)\,:\,$([g^0],[g^2],[g^2])$ 
\end{tabular}
&
$\begin{pmatrix}
* & 0 & 0 \\
0 & * & * \\
0 & * & *
\end{pmatrix}$ 
& 
$\begin{pmatrix}
* & * & * \\
* & * & * \\
* & * & *
\end{pmatrix}$
&
$\begin{pmatrix}
* & 0 & 0 \\
0 & * & * \\
0 & * & *
\end{pmatrix}$\\ \hline

\begin{tabular}{l}
(vi)\,:\,$([g^1],[g^1],[g^2])$ 
\end{tabular}
&
$\begin{pmatrix}
* & * & 0 \\
* & * & 0 \\
0 & 0 & *
\end{pmatrix}$ 
& 
$\begin{pmatrix}
* & * & * \\
* & * & * \\
* & * & *
\end{pmatrix}$
&
$\begin{pmatrix}
* & * & * \\
* & * & * \\
* & * & *
\end{pmatrix}$\\ \hline

\begin{tabular}{l}
(vii)\,:\,$([g^1],[g^2],[g^2])$ 
\end{tabular}
&
$\begin{pmatrix}
* & 0 & 0 \\
0 & * & * \\
0 & * & *
\end{pmatrix}$ 
& 
$\begin{pmatrix}
* & * & * \\
* & * & * \\
* & * & *
\end{pmatrix}$
&
$\begin{pmatrix}
* & * & * \\
* & * & * \\
* & * & *
\end{pmatrix}$\\ \hline
    \end{tabular}
    \caption{${C_{\mathrm{W}}}^{ij}$ matrices for $M=5$.   The asterisk symbols "$*$" denote non-vanishing elements.}
    \label{tab:C-M=5}
\end{table}

\subsection{Combinations of charged lepton mass matrix and neutrino mass matrix}
\label{sec:Y-C}

Here, we study the Yukawa couplings of the charged lepton sector, 
\begin{align}
    Y^{ij}\bar L_iHe_j,
\end{align}
where $e_j$ denote right-handed charged leptons, 
and combine their textures with neutrino mass textures, which are obtained in the previous subsection.
Analysis on $Y^{ij}$ is similar to one in the quark sector \cite{Kobayashi:2024cvp,Kobayashi:2025znw}.

\subsubsection{$M=3$}
\label{sec:Y-C-M=3}

We study the case with $M=3$.
We use the same flavor assignments for $e_i$ as ones for $L_i$ in Eq.~(\ref{eq:flavor-assign-M=3}).
The coupling selection rules discussed in section \ref{sec:Z2-gauging} forbids the $\phi_0 \phi_0 \phi_1$ coupling, where $\phi_0$ and $\phi_1$ correspond to $[g^0]$ and $[g^1]$, and allows the other 3-point couplings.
By use of these selection rules, we can examine the possible textures of $Y^{ij}$.
Table \ref{tab:Y-C-M=3} shows possible combinations of $Y^{ij}$ and ${C_{\mathrm{W}}}^{ij}$.
Although all textures of ${C_{\mathrm{W}}}^{ij}$ can be understood effectively by the $\mathbb{Z}_2$ symmetry, 
all textures of $Y^{ij}$ can not be obtained by the $\mathbb{Z}_2$ symmetry.

In the Standard Model, the Higgs mode is single, and it must correspond to a single class, $[g^0]$ or $[g^1]$ for both $Y^{ij}$ and ${C_{\mathrm{W}}}^{ij}$.
On the other hand, the Higgs mode $H_u$ appearing in the Weinberg operators  is different from the Higgs mode $H_d$ appearing in the Yukawa couplings of charged leptons e.g., in supersymmetric standard models\footnote{If $H_u$ and $H_d$ correspond to different classes, the $\mu$-term $\mu H_u H_d$ is not allowed in supersymmetric standard model. We have to introduce a singlet $S$ with a proper class like the next-to minimal supersymmetric standard model to generate the term $\lambda SH_u H_d$.} 
and type II non-supersymmetric two Higgs doublet models. 
In general, these $H_u$ and $H_d$ modes can correspond to 
different classes each other.
For example, for the first flavor combination (i,i), we can obtain 
\begin{align}
    Y^{ij}=
    \begin{pmatrix}
        0 & * & * \\
        * & * & * \\
        * & * & *
    \end{pmatrix}, \qquad 
    {C_{\mathrm{W}}}^{ij}=
    \begin{pmatrix}
        * & 0 & 0 \\
        0 & * & * \\
        0 & * & *
    \end{pmatrix},
\end{align}
when $H_u$ and $H_d$ correspond to $[g^0]$ and $[g^1]$, respectively.

\begin{table}[H]
    \centering
    \begin{tabular}{|c|c|c|c|} \hline
        Flavor & Higgs\ $[g^0]$ & Higgs\ $[g^1]$  \\ 
        (Left,Right) & $Y^{ij}$,\hspace{1cm} ${C_{\mathrm{W}}}^{ij}$ & $Y^{ij}$,\hspace{1cm} ${C_{\mathrm{W}}}^{ij}$\\ \hline
(\text{i},\text{i}) 
&  
$\begin{pmatrix}
* & 0 & 0 \\
0 & * & * \\
0 & * & *
\end{pmatrix}$

$\begin{pmatrix}
* & 0 & 0 \\
0 & * & * \\
0 & * & *
\end{pmatrix}$
&
$\begin{pmatrix}
0 & * & * \\
* & * & * \\
* & * & *
\end{pmatrix}$

$\begin{pmatrix}
* & * & * \\
* & * & * \\
* & * & *
\end{pmatrix}$
\\ \hline

(\text{i},\text{ii}) 
&  
$\begin{pmatrix}
* & * & 0 \\
0 & 0 & * \\
0 & 0 & *
\end{pmatrix}$

$\begin{pmatrix}
* & 0 & 0 \\
0 & * & * \\
0 & * & *
\end{pmatrix}$
&
$\begin{pmatrix}
0 & 0 & * \\
* & * & * \\
* & * & *
\end{pmatrix}$

$\begin{pmatrix}
* & * & * \\
* & * & * \\
* & * & *
\end{pmatrix}$
\\ \hline

(\text{ii},\text{i}) 
&  
$\begin{pmatrix}
* & 0 & 0 \\
* & 0 & 0 \\
0 & * & *
\end{pmatrix}$

$\begin{pmatrix}
* & * & 0 \\
* & * & 0 \\
0 & 0 & *
\end{pmatrix}$
&
$\begin{pmatrix}
0 & * & * \\
0 & * & * \\
* & * & *
\end{pmatrix}$

$\begin{pmatrix}
* & * & * \\
* & * & * \\
* & * & *
\end{pmatrix}$
\\ \hline

(\text{ii},\text{ii}) 
&  
$\begin{pmatrix}
* & * & 0 \\
* & * & 0 \\
0 & 0 & *
\end{pmatrix}$

$\begin{pmatrix}
* & * & 0 \\
* & * & 0 \\
0 & 0 & *
\end{pmatrix}$
&
$\begin{pmatrix}
0 & 0 & * \\
0 & 0 & * \\
* & * & *
\end{pmatrix}$

$\begin{pmatrix}
* & * & * \\
* & * & * \\
* & * & *
\end{pmatrix}$
\\ \hline

    \end{tabular}
    \caption{$Y^{ij}$ and ${C_{\mathrm{W}}}^{ij}$ matrices for $M=3$.}
    \label{tab:Y-C-M=3}
\end{table}

\subsubsection{$M=4$}
\label{sec:Y-C-M=4}

We study the case with $M=4$.
We use the same flavor assignments for $e_i$ as ones for $L_i$ in Table~\ref{tab:Flavor-M=4}.
The coupling selection rules discussed in section \ref{sec:Z2-gauging} allow the following couplings:
\begin{align}
    \phi_0 \phi_0 \phi_0, \qquad \phi_0 \phi_1 \phi_1, \qquad \phi_0 \phi_2 \phi_2 ,\qquad \phi_1 \phi_1 \phi_2,
\end{align}
where $\phi_k$ corresponds to the class $[g^k]$.
By use of these selection rules, we can examine the possible textures of $Y^{ij}$.
Tables~\ref{tab:Y-C-M=4-1},\,\ref{tab:Y-C-M=4-2},\,\ref{tab:Y-C-M=4-3},\,\ref{tab:Y-C-M=4-4},\,\ref{tab:Y-C-M=4-5} show possible combinations of $Y^{ij}$ and ${C_{\mathrm{W}}}^{ij}$. 
In the Standard Model, the Higgs mode corresponds to a single class.
However, $H_u$ and $H_d$ can correspond to different classes in supersymmetric standard models and two Higgs doublet models.

\newpage

\begin{table}[H]
    \centering
    \begin{tabular}{|c|c|c|c|} \hline
        Flavor & Higgs\ $[g^0]$ & Higgs\ $[g^1]$ &Higgs\ $[g^2]$ \\ 
        (Left,Right) & $Y^{ij}$,\hspace{1cm} ${C_{\mathrm{W}}}^{ij}$ & $Y^{ij}$,\hspace{1cm} ${C_{\mathrm{W}}}^{ij}$ & $Y^{ij}$,\hspace{1cm} ${C_{\mathrm{W}}}^{ij}$\\\hline
(\text{i},\text{i}) 
&  
$\begin{pmatrix}
* & 0 & 0 \\
0 & * & 0 \\
0 & 0 & *
\end{pmatrix}$

$\begin{pmatrix}
* & 0 & 0 \\
0 & * & 0 \\
0 & 0 & *
\end{pmatrix}$
&
$\begin{pmatrix}
0 & * & 0 \\
* & 0 & * \\
0 & * & 0
\end{pmatrix}$

$\begin{pmatrix}
* & 0 & * \\
0 & * & 0 \\
* & 0 & *
\end{pmatrix}$
&
$\begin{pmatrix}
0 & 0 & * \\
0 & * & 0 \\
* & 0 & 0
\end{pmatrix}$

$\begin{pmatrix}
* & 0 & 0 \\
0 & * & 0 \\
0 & 0 & *
\end{pmatrix}$
\\ \hline

(\text{i},\text{ii})  & $\begin{pmatrix}
* & * & 0 \\
0 & 0 & * \\
0 & 0 & 0
\end{pmatrix}$

$\begin{pmatrix}
* & 0 & 0 \\
0 & * & 0 \\
0 & 0 & *
\end{pmatrix}$
&
$\begin{pmatrix}
0 & 0 & * \\
* & * & 0 \\
0 & 0 & *
\end{pmatrix}$

$\begin{pmatrix}
* & 0 & * \\
0 & * & 0 \\
* & 0 & *
\end{pmatrix}$
&
$\begin{pmatrix}
0 & 0 & 0 \\
0 & 0 & * \\
* & * & 0
\end{pmatrix}$

$\begin{pmatrix}
* & 0 & 0 \\
0 & * & 0 \\
0 & 0 & *
\end{pmatrix}$
\\ \hline

(\text{i},\text{iii})  & $\begin{pmatrix}
* & * & 0 \\
0 & 0 & 0 \\
0 & 0 & *
\end{pmatrix}$

$\begin{pmatrix}
* & 0 & 0 \\
0 & * & 0 \\
0 & 0 & *
\end{pmatrix}$
&
$\begin{pmatrix}
0 & 0 & 0 \\
* & * & * \\
0 & 0 & 0
\end{pmatrix}$

$\begin{pmatrix}
* & 0 & * \\
0 & * & 0 \\
* & 0 & *
\end{pmatrix}$
&
$\begin{pmatrix}
0 & 0 & * \\
0 & 0 & 0 \\
* & * & 0
\end{pmatrix}$

$\begin{pmatrix}
* & 0 & 0 \\
0 & * & 0 \\
0 & 0 & *
\end{pmatrix}$
\\ \hline

(\text{i},\text{iv})  & $\begin{pmatrix}
* & 0 & 0 \\
0 & * & * \\
0 & 0 & 0
\end{pmatrix}$

$\begin{pmatrix}
* & 0 & 0 \\
0 & * & 0 \\
0 & 0 & *
\end{pmatrix}$
&
$\begin{pmatrix}
0 & * & * \\
* & 0 & 0 \\
0 & * & *
\end{pmatrix}$

$\begin{pmatrix}
* & 0 & * \\
0 & * & 0 \\
* & 0 & *
\end{pmatrix}$
&
$\begin{pmatrix}
0 & 0 & 0 \\
0 & * & * \\
* & 0 & 0
\end{pmatrix}$

$\begin{pmatrix}
* & 0 & 0 \\
0 & * & 0 \\
0 & 0 & *
\end{pmatrix}$
\\ \hline

(\text{i},\text{v})  & $\begin{pmatrix}
* & 0 & 0 \\
0 & 0 & 0 \\
0 & * & *
\end{pmatrix}$

$\begin{pmatrix}
* & 0 & 0 \\
0 & * & 0 \\
0 & 0 & *
\end{pmatrix}$
&
$\begin{pmatrix}
0 & 0 & 0 \\
* & * & * \\
0 & 0 & 0
\end{pmatrix}$

$\begin{pmatrix}
* & 0 & * \\
0 & * & 0 \\
* & 0 & *
\end{pmatrix}$
&
$\begin{pmatrix}
0 & * & * \\
0 & 0 & 0 \\
* & 0 & 0
\end{pmatrix}$

$\begin{pmatrix}
* & 0 & 0 \\
0 & * & 0 \\
0 & 0 & *
\end{pmatrix}$
\\ \hline

(\text{i},\text{vi})   & $\begin{pmatrix}
0 & 0 & 0 \\
* & * & 0 \\
0 & 0 & *
\end{pmatrix}$

$\begin{pmatrix}
* & 0 & 0 \\
0 & * & 0 \\
0 & 0 & *
\end{pmatrix}$
&
$\begin{pmatrix}
* & * & 0 \\
0 & 0 & * \\
* & * & 0
\end{pmatrix}$

$\begin{pmatrix}
* & 0 & * \\
0 & * & 0 \\
* & 0 & *
\end{pmatrix}$
&
$\begin{pmatrix}
0 & 0 & * \\
* & * & 0 \\
0 & 0 & 0
\end{pmatrix}$

$\begin{pmatrix}
* & 0 & 0 \\
0 & * & 0 \\
0 & 0 & *
\end{pmatrix}$
\\ \hline

(\text{i},\text{vii})  & $\begin{pmatrix}
0 & 0 & 0 \\
* & 0 & 0 \\
0 & * & *
\end{pmatrix}$

$\begin{pmatrix}
* & 0 & 0 \\
0 & * & 0 \\
0 & 0 & *
\end{pmatrix}$
&
$\begin{pmatrix}
* & 0 & 0 \\
0 & * & * \\
* & 0 & 0
\end{pmatrix}$

$\begin{pmatrix}
* & 0 & * \\
0 & * & 0 \\
* & 0 & *
\end{pmatrix}$
&
$\begin{pmatrix}
0 & * & * \\
* & 0 & 0 \\
0 & 0 & 0
\end{pmatrix}$

$\begin{pmatrix}
* & 0 & 0 \\
0 & * & 0 \\
0 & 0 & *
\end{pmatrix}$
\\ \hline

(\text{ii},\text{i})  & $\begin{pmatrix}
* & 0 & 0 \\
* & 0 & 0 \\
0 & * & 0
\end{pmatrix}$

$\begin{pmatrix}
* & * & 0 \\
* & * & 0 \\
0 & 0 & *
\end{pmatrix}$
&
$\begin{pmatrix}
0 & * & 0 \\
0 & * & 0 \\
* & 0 & *
\end{pmatrix}$

$\begin{pmatrix}
* & * & 0 \\
* & * & 0 \\
0 & 0 & *
\end{pmatrix}$
&
$\begin{pmatrix}
0 & 0 & * \\
0 & 0 & * \\
0 & * & 0
\end{pmatrix}$

$\begin{pmatrix}
* & * & 0 \\
* & * & 0 \\
0 & 0 & *
\end{pmatrix}$
\\ \hline

(\text{ii},\text{ii})  & $\begin{pmatrix}
* & * & 0 \\
* & * & 0 \\
0 & 0 & *
\end{pmatrix}$

$\begin{pmatrix}
* & * & 0 \\
* & * & 0 \\
0 & 0 & *
\end{pmatrix}$
&
$\begin{pmatrix}
0 & 0 & * \\
0 & 0 & * \\
* & * & 0
\end{pmatrix}$

$\begin{pmatrix}
* & * & 0 \\
* & * & 0 \\
0 & 0 & *
\end{pmatrix}$
&
$\begin{pmatrix}
0 & 0 & 0 \\
0 & 0 & 0 \\
0 & 0 & *
\end{pmatrix}$

$\begin{pmatrix}
* & * & 0 \\
* & * & 0 \\
0 & 0 & *
\end{pmatrix}$
\\ \hline

(\text{ii},\text{iii})   
& 
$\begin{pmatrix}
* & * & 0 \\
* & * & 0 \\
0 & 0 & 0
\end{pmatrix}$

$\begin{pmatrix}
* & * & 0 \\
* & * & 0 \\
0 & 0 & *
\end{pmatrix}$
&
$\begin{pmatrix}
0 & 0 & 0 \\
0 & 0 & 0 \\
* & * & *
\end{pmatrix}$

$\begin{pmatrix}
* & * & 0 \\
* & * & 0 \\
0 & 0 & *
\end{pmatrix}$
&
$\begin{pmatrix}
0 & 0 & * \\
0 & 0 & * \\
0 & 0 & 0
\end{pmatrix}$

$\begin{pmatrix}
* & * & 0 \\
* & * & 0 \\
0 & 0 & *
\end{pmatrix}$
\\ \hline

(\text{ii},\text{iv})   
&
$\begin{pmatrix}
* & 0 & 0 \\
* & 0 & 0 \\
0 & * & *
\end{pmatrix}$

$\begin{pmatrix}
* & * & 0 \\
* & * & 0 \\
0 & 0 & *
\end{pmatrix}$
&
$\begin{pmatrix}
0 & * & * \\
0 & * & * \\
* & 0 & 0
\end{pmatrix}$

$\begin{pmatrix}
* & * & 0 \\
* & * & 0 \\
0 & 0 & *
\end{pmatrix}$
&
$\begin{pmatrix}
0 & 0 & 0 \\
0 & 0 & 0 \\
0 & * & *
\end{pmatrix}$

$\begin{pmatrix}
* & * & 0 \\
* & * & 0 \\
0 & 0 & *
\end{pmatrix}$
\\ \hline

(\text{ii},\text{v})   & $\begin{pmatrix}
* & 0 & 0 \\
* & 0 & 0 \\
0 & 0 & 0
\end{pmatrix}$

$\begin{pmatrix}
* & * & 0 \\
* & * & 0 \\
0 & 0 & *
\end{pmatrix}$
&
$\begin{pmatrix}
0 & 0 & 0 \\
0 & 0 & 0 \\
* & * & *
\end{pmatrix}$

$\begin{pmatrix}
* & * & 0 \\
* & * & 0 \\
0 & 0 & *
\end{pmatrix}$
&
$\begin{pmatrix}
0 & * & * \\
0 & * & * \\
0 & 0 & 0
\end{pmatrix}$

$\begin{pmatrix}
* & * & 0 \\
* & * & 0 \\
0 & 0 & *
\end{pmatrix}$
\\ \hline
    \end{tabular}
    \caption{$Y^{ij}$ and ${C_{\mathrm{W}}}^{ij}$ matrices for $M=4$.}
    \label{tab:Y-C-M=4-1}
\end{table}

\newpage

\begin{table}[H]
    \centering
    \begin{tabular}{|c|c|c|c|} \hline
        Flavor & Higgs\ $[g^0]$ & Higgs\ $[g^1]$ &Higgs\ $[g^2]$ \\ 
        (Left,Right) & $Y^{ij}$,\hspace{1cm} ${C_{\mathrm{W}}}^{ij}$ & $Y^{ij}$,\hspace{1cm} ${C_{\mathrm{W}}}^{ij}$ & $Y^{ij}$,\hspace{1cm} ${C_{\mathrm{W}}}^{ij}$\\\hline

(\text{ii},\text{vi}) & $\begin{pmatrix}
0 & 0 & 0 \\
0 & 0 & 0 \\
* & * & 0
\end{pmatrix}$

$\begin{pmatrix}
* & * & 0 \\
* & * & 0 \\
0 & 0 & *
\end{pmatrix}$
&
$\begin{pmatrix}
* & * & 0 \\
* & * & 0 \\
0 & 0 & *
\end{pmatrix}$

$\begin{pmatrix}
* & * & 0 \\
* & * & 0 \\
0 & 0 & *
\end{pmatrix}$
&
$\begin{pmatrix}
0 & 0 & * \\
0 & 0 & * \\
* & * & 0
\end{pmatrix}$

$\begin{pmatrix}
* & * & 0 \\
* & * & 0 \\
0 & 0 & *
\end{pmatrix}$
\\ \hline

(\text{ii},\text{vii}) &  $\begin{pmatrix}
0 & 0 & 0 \\
0 & 0 & 0 \\
* & 0 & 0
\end{pmatrix}$

$\begin{pmatrix}
* & * & 0 \\
* & * & 0 \\
0 & 0 & *
\end{pmatrix}$
&
$\begin{pmatrix}
* & 0 & 0 \\
* & 0 & 0 \\
0 & * & *
\end{pmatrix}$

$\begin{pmatrix}
* & * & 0 \\
* & * & 0 \\
0 & 0 & *
\end{pmatrix}$
&
$\begin{pmatrix}
0 & * & * \\
0 & * & * \\
* & 0 & 0
\end{pmatrix}$

$\begin{pmatrix}
* & * & 0 \\
* & * & 0 \\
0 & 0 & *
\end{pmatrix}$
\\ \hline

(\text{iii},\text{i})  & $\begin{pmatrix}
* & 0 & 0 \\
* & 0 & 0 \\
0 & 0 & *
\end{pmatrix}$

$\begin{pmatrix}
* & * & 0 \\
* & * & 0 \\
0 & 0 & *
\end{pmatrix}$
&
$\begin{pmatrix}
0 & * & 0 \\
0 & * & 0 \\
0 & * & 0
\end{pmatrix}$

$\begin{pmatrix}
* & * & * \\
* & * & * \\
* & * & *
\end{pmatrix}$
&
$\begin{pmatrix}
0 & 0 & * \\
0 & 0 & * \\
* & 0 & 0
\end{pmatrix}$

$\begin{pmatrix}
* & * & 0 \\
* & * & 0 \\
0 & 0 & *
\end{pmatrix}$
\\ \hline

(\text{iii},\text{ii})   
& 
$\begin{pmatrix}
* & * & 0 \\
* & * & 0 \\
0 & 0 & 0
\end{pmatrix}$

$\begin{pmatrix}
* & * & 0 \\
* & * & 0 \\
0 & 0 & *
\end{pmatrix}$
&
$\begin{pmatrix}
0 & 0 & * \\
0 & 0 & * \\
0 & 0 & *
\end{pmatrix}$

$\begin{pmatrix}
* & * & * \\
* & * & * \\
* & * & *
\end{pmatrix}$
&
$\begin{pmatrix}
0 & 0 & 0 \\
0 & 0 & 0 \\
* & * & 0
\end{pmatrix}$

$\begin{pmatrix}
* & * & 0 \\
* & * & 0 \\
0 & 0 & *
\end{pmatrix}$
\\ \hline

(\text{iii},\text{iii})  & $\begin{pmatrix}
* & * & 0 \\
* & * & 0 \\
0 & 0 & *
\end{pmatrix}$

$\begin{pmatrix}
* & * & 0 \\
* & * & 0 \\
0 & 0 & *
\end{pmatrix}$
&
$\begin{pmatrix}
0 & 0 & 0 \\
0 & 0 & 0 \\
0 & 0 & 0
\end{pmatrix}$

$\begin{pmatrix}
* & * & * \\
* & * & * \\
* & * & *
\end{pmatrix}$
&
$\begin{pmatrix}
0 & 0 & * \\
0 & 0 & * \\
* & * & 0
\end{pmatrix}$

$\begin{pmatrix}
* & * & 0 \\
* & * & 0 \\
0 & 0 & *
\end{pmatrix}$
\\ \hline

(\text{iii},\text{iv})  & $\begin{pmatrix}
* & 0 & 0 \\
* & 0 & 0 \\
0 & 0 & 0
\end{pmatrix}$

$\begin{pmatrix}
* & * & 0 \\
* & * & 0 \\
0 & 0 & *
\end{pmatrix}$
&
$\begin{pmatrix}
0 & * & * \\
0 & * & * \\
0 & * & *
\end{pmatrix}$

$\begin{pmatrix}
* & * & * \\
* & * & * \\
* & * & *
\end{pmatrix}$
&
$\begin{pmatrix}
0 & 0 & 0 \\
0 & 0 & 0 \\
* & 0 & 0
\end{pmatrix}$

$\begin{pmatrix}
* & * & 0 \\
* & * & 0 \\
0 & 0 & *
\end{pmatrix}$
\\ \hline

(\text{iii},\text{v})  & $\begin{pmatrix}
* & 0 & 0 \\
* & 0 & 0 \\
0 & * & *
\end{pmatrix}$

$\begin{pmatrix}
* & * & 0 \\
* & * & 0 \\
0 & 0 & *
\end{pmatrix}$
&
$\begin{pmatrix}
0 & 0 & 0 \\
0 & 0 & 0 \\
0 & 0 & 0
\end{pmatrix}$

$\begin{pmatrix}
* & * & * \\
* & * & * \\
* & * & *
\end{pmatrix}$
&
$\begin{pmatrix}
0 & * & * \\
0 & * & * \\
* & 0 & 0
\end{pmatrix}$

$\begin{pmatrix}
* & * & 0 \\
* & * & 0 \\
0 & 0 & *
\end{pmatrix}$
\\ \hline

(\text{iii},\text{vi})  & $\begin{pmatrix}
0 & 0 & 0 \\
0 & 0 & 0 \\
0 & 0 & *
\end{pmatrix}$

$\begin{pmatrix}
* & * & 0 \\
* & * & 0 \\
0 & 0 & *
\end{pmatrix}$
&
$\begin{pmatrix}
* & * & 0 \\
* & * & 0 \\
* & * & 0
\end{pmatrix}$

$\begin{pmatrix}
* & * & * \\
* & * & * \\
* & * & *
\end{pmatrix}$
&
$\begin{pmatrix}
0 & 0 & * \\
0 & 0 & * \\
0 & 0 & 0
\end{pmatrix}$

$\begin{pmatrix}
* & * & 0 \\
* & * & 0 \\
0 & 0 & *
\end{pmatrix}$
\\ \hline

(\text{iii},\text{vii})   & $\begin{pmatrix}
0 & 0 & 0 \\
0 & 0 & 0 \\
0 & * & *
\end{pmatrix}$

$\begin{pmatrix}
* & * & 0 \\
* & * & 0 \\
0 & 0 & *
\end{pmatrix}$
&
$\begin{pmatrix}
* & 0 & 0 \\
* & 0 & 0 \\
* & 0 & 0
\end{pmatrix}$

$\begin{pmatrix}
* & * & * \\
* & * & * \\
* & * & *
\end{pmatrix}$
&
$\begin{pmatrix}
0 & * & * \\
0 & * & * \\
0 & 0 & 0
\end{pmatrix}$

$\begin{pmatrix}
* & * & 0 \\
* & * & 0 \\
0 & 0 & *
\end{pmatrix}$
\\ \hline

(\text{iv},\text{i})  & $\begin{pmatrix}
* & 0 & 0 \\
0 & * & 0 \\
0 & * & 0
\end{pmatrix}$

$\begin{pmatrix}
* & 0 & 0 \\
0 & * & * \\
0 & * & *
\end{pmatrix}$
&
$\begin{pmatrix}
0 & * & 0 \\
* & 0 & * \\
* & 0 & *
\end{pmatrix}$

$\begin{pmatrix}
* & 0 & 0 \\
0 & * & * \\
0 & * & *
\end{pmatrix}$
&
$\begin{pmatrix}
0 & 0 & * \\
0 & * & 0 \\
0 & * & 0
\end{pmatrix}$

$\begin{pmatrix}
* & 0 & 0 \\
0 & * & * \\
0 & * & *
\end{pmatrix}$
\\ \hline

(\text{iv},\text{ii})   
&
$\begin{pmatrix}
* & * & 0 \\
0 & 0 & * \\
0 & 0 & *
\end{pmatrix}$

$\begin{pmatrix}
* & 0 & 0 \\
0 & * & * \\
0 & * & *
\end{pmatrix}$
&
$\begin{pmatrix}
0 & 0 & * \\
* & * & 0 \\
* & * & 0
\end{pmatrix}$

$\begin{pmatrix}
* & 0 & 0 \\
0 & * & * \\
0 & * & *
\end{pmatrix}$
&
$\begin{pmatrix}
0 & 0 & 0 \\
0 & 0 & * \\
0 & 0 & *
\end{pmatrix}$

$\begin{pmatrix}
* & 0 & 0 \\
0 & * & * \\
0 & * & *
\end{pmatrix}$
\\ \hline

(\text{iv},\text{iii})  & $\begin{pmatrix}
* & * & 0 \\
0 & 0 & 0 \\
0 & 0 & 0
\end{pmatrix}$

$\begin{pmatrix}
* & 0 & 0 \\
0 & * & * \\
0 & * & *
\end{pmatrix}$
&
$\begin{pmatrix}
0 & 0 & 0 \\
* & * & * \\
* & * & *
\end{pmatrix}$

$\begin{pmatrix}
* & 0 & 0 \\
0 & * & * \\
0 & * & *
\end{pmatrix}$
&
$\begin{pmatrix}
0 & 0 & * \\
0 & 0 & 0 \\
0 & 0 & 0
\end{pmatrix}$

$\begin{pmatrix}
* & 0 & 0 \\
0 & * & * \\
0 & * & *
\end{pmatrix}$
\\ \hline
    \end{tabular}
    \caption{$Y^{ij}$ and ${C_{\mathrm{W}}}^{ij}$ matrices for $M=4$.}
    \label{tab:Y-C-M=4-2}
\end{table}

\newpage

\begin{table}[H]
    \centering
    \begin{tabular}{|c|c|c|c|} \hline
        Flavor & Higgs\ $[g^0]$ & Higgs\ $[g^1]$ &Higgs\ $[g^2]$ \\ 
        (Left,Right) & $Y^{ij}$,\hspace{1cm} ${C_{\mathrm{W}}}^{ij}$ & $Y^{ij}$,\hspace{1cm} ${C_{\mathrm{W}}}^{ij}$ & $Y^{ij}$,\hspace{1cm} ${C_{\mathrm{W}}}^{ij}$\\\hline
        
(\text{iv},\text{iv}) 
& 
$\begin{pmatrix}
* & 0 & 0 \\
0 & * & * \\
0 & * & *
\end{pmatrix}$

$\begin{pmatrix}
* & 0 & 0 \\
0 & * & * \\
0 & * & *
\end{pmatrix}$
&
$\begin{pmatrix}
0 & * & * \\
* & 0 & 0 \\
* & 0 & 0
\end{pmatrix}$

$\begin{pmatrix}
* & 0 & 0 \\
0 & * & * \\
0 & * & *
\end{pmatrix}$
&
$\begin{pmatrix}
0 & 0 & 0 \\
0 & * & * \\
0 & * & *
\end{pmatrix}$

$\begin{pmatrix}
* & 0 & 0 \\
0 & * & * \\
0 & * & *
\end{pmatrix}$
\\ \hline

(\text{iv},\text{v})  & $\begin{pmatrix}
* & 0 & 0 \\
0 & 0 & 0 \\
0 & 0 & 0
\end{pmatrix}$

$\begin{pmatrix}
* & 0 & 0 \\
0 & * & * \\
0 & * & *
\end{pmatrix}$
&
$\begin{pmatrix}
0 & 0 & 0 \\
* & * & * \\
* & * & *
\end{pmatrix}$

$\begin{pmatrix}
* & 0 & 0 \\
0 & * & * \\
0 & * & *
\end{pmatrix}$
&
$\begin{pmatrix}
0 & * & * \\
0 & 0 & 0 \\
0 & 0 & 0
\end{pmatrix}$

$\begin{pmatrix}
* & 0 & 0 \\
0 & * & * \\
0 & * & *
\end{pmatrix}$
\\ \hline

(\text{iv},\text{vi})   & $\begin{pmatrix}
0 & 0 & 0 \\
* & * & 0 \\
* & * & 0
\end{pmatrix}$

$\begin{pmatrix}
* & 0 & 0 \\
0 & * & * \\
0 & * & *
\end{pmatrix}$
&
$\begin{pmatrix}
* & * & 0 \\
0 & 0 & * \\
0 & 0 & *
\end{pmatrix}$

$\begin{pmatrix}
* & 0 & 0 \\
0 & * & * \\
0 & * & *
\end{pmatrix}$
&
$\begin{pmatrix}
0 & 0 & * \\
* & * & 0 \\
* & * & 0
\end{pmatrix}$

$\begin{pmatrix}
* & 0 & 0 \\
0 & * & * \\
0 & * & *
\end{pmatrix}$
\\ \hline

(\text{iv},\text{vii})   & $\begin{pmatrix}
0 & 0 & 0 \\
* & 0 & 0 \\
* & 0 & 0
\end{pmatrix}$

$\begin{pmatrix}
* & 0 & 0 \\
0 & * & * \\
0 & * & *
\end{pmatrix}$
&
$\begin{pmatrix}
* & 0 & 0 \\
0 & * & * \\
0 & * & *
\end{pmatrix}$

$\begin{pmatrix}
* & 0 & 0 \\
0 & * & * \\
0 & * & *
\end{pmatrix}$
&
$\begin{pmatrix}
0 & * & * \\
* & 0 & 0 \\
* & 0 & 0
\end{pmatrix}$

$\begin{pmatrix}
* & 0 & 0 \\
0 & * & * \\
0 & * & *
\end{pmatrix}$
\\ \hline

(\text{v},\text{i})  & $\begin{pmatrix}
* & 0 & 0 \\
0 & 0 & * \\
0 & 0 & *
\end{pmatrix}$

$\begin{pmatrix}
* & 0 & 0 \\
0 & * & * \\
0 & * & *
\end{pmatrix}$
&
$\begin{pmatrix}
0 & * & 0 \\
0 & * & 0 \\
0 & * & 0
\end{pmatrix}$

$\begin{pmatrix}
* & * & * \\
* & * & * \\
* & * & *
\end{pmatrix}$
&
$\begin{pmatrix}
0 & 0 & * \\
* & 0 & 0 \\
* & 0 & 0
\end{pmatrix}$

$\begin{pmatrix}
* & 0 & 0 \\
0 & * & * \\
0 & * & *
\end{pmatrix}$
\\ \hline

(\text{v},\text{ii})  
&
$\begin{pmatrix}
* & * & 0 \\
0 & 0 & 0 \\
0 & 0 & 0
\end{pmatrix}$

$\begin{pmatrix}
* & 0 & 0 \\
0 & * & * \\
0 & * & *
\end{pmatrix}$
&
$\begin{pmatrix}
0 & 0 & * \\
0 & 0 & * \\
0 & 0 & *
\end{pmatrix}$

$\begin{pmatrix}
* & * & * \\
* & * & * \\
* & * & *
\end{pmatrix}$
&
$\begin{pmatrix}
0 & 0 & 0 \\
* & * & 0 \\
* & * & 0
\end{pmatrix}$

$\begin{pmatrix}
* & 0 & 0 \\
0 & * & * \\
0 & * & *
\end{pmatrix}$
\\ \hline

(\text{v},\text{iii})  
&
$\begin{pmatrix}
* & * & 0 \\
0 & 0 & * \\
0 & 0 & *
\end{pmatrix}$

$\begin{pmatrix}
* & 0 & 0 \\
0 & * & * \\
0 & * & *
\end{pmatrix}$
&
$\begin{pmatrix}
0 & 0 & 0 \\
0 & 0 & 0 \\
0 & 0 & 0
\end{pmatrix}$

$\begin{pmatrix}
* & * & * \\
* & * & * \\
* & * & *
\end{pmatrix}$
&
$\begin{pmatrix}
0 & 0 & * \\
* & * & 0 \\
* & * & 0
\end{pmatrix}$

$\begin{pmatrix}
* & 0 & 0 \\
0 & * & * \\
0 & * & *
\end{pmatrix}$
\\ \hline

(\text{v},\text{iv})  & $\begin{pmatrix}
* & 0 & 0 \\
0 & 0 & 0 \\
0 & 0 & 0
\end{pmatrix}$

$\begin{pmatrix}
* & 0 & 0 \\
0 & * & * \\
0 & * & *
\end{pmatrix}$
&
$\begin{pmatrix}
0 & * & * \\
0 & * & * \\
0 & * & *
\end{pmatrix}$

$\begin{pmatrix}
* & * & * \\
* & * & * \\
* & * & *
\end{pmatrix}$
&
$\begin{pmatrix}
0 & 0 & 0 \\
* & 0 & 0 \\
* & 0 & 0
\end{pmatrix}$

$\begin{pmatrix}
* & 0 & 0 \\
0 & * & * \\
0 & * & *
\end{pmatrix}$
\\ \hline

(\text{v},\text{v})   & $\begin{pmatrix}
* & 0 & 0 \\
0 & * & * \\
0 & * & *
\end{pmatrix}$

$\begin{pmatrix}
* & 0 & 0 \\
0 & * & * \\
0 & * & *
\end{pmatrix}$
&
$\begin{pmatrix}
0 & 0 & 0 \\
0 & 0 & 0 \\
0 & 0 & 0
\end{pmatrix}$

$\begin{pmatrix}
* & * & * \\
* & * & * \\
* & * & *
\end{pmatrix}$
&
$\begin{pmatrix}
0 & * & * \\
* & 0 & 0 \\
* & 0 & 0
\end{pmatrix}$

$\begin{pmatrix}
* & 0 & 0 \\
0 & * & * \\
0 & * & *
\end{pmatrix}$
\\ \hline

(\text{v},\text{vi}) 
&
$\begin{pmatrix}
0 & 0 & 0 \\
0 & 0 & * \\
0 & 0 & *
\end{pmatrix}$

$\begin{pmatrix}
* & 0 & 0 \\
0 & * & * \\
0 & * & *
\end{pmatrix}$
&
$\begin{pmatrix}
* & * & 0 \\
* & * & 0 \\
* & * & 0
\end{pmatrix}$

$\begin{pmatrix}
* & * & * \\
* & * & * \\
* & * & *
\end{pmatrix}$
&
$\begin{pmatrix}
0 & 0 & * \\
0 & 0 & 0 \\
0 & 0 & 0
\end{pmatrix}$

$\begin{pmatrix}
* & 0 & 0 \\
0 & * & * \\
0 & * & *
\end{pmatrix}$
\\ \hline

(\text{v},\text{vii}) 
& 
$\begin{pmatrix}
0 & 0 & 0 \\
0 & * & * \\
0 & * & *
\end{pmatrix}$

$\begin{pmatrix}
* & 0 & 0 \\
0 & * & * \\
0 & * & *
\end{pmatrix}$
&
$\begin{pmatrix}
* & 0 & 0 \\
* & 0 & 0 \\
* & 0 & 0
\end{pmatrix}$

$\begin{pmatrix}
* & * & * \\
* & * & * \\
* & * & *
\end{pmatrix}$
&
$\begin{pmatrix}
0 & * & * \\
0 & 0 & 0 \\
0 & 0 & 0
\end{pmatrix}$

$\begin{pmatrix}
* & 0 & 0 \\
0 & * & * \\
0 & * & *
\end{pmatrix}$
\\ \hline

(\text{vi},\text{i})   & $\begin{pmatrix}
0 & * & 0 \\
0 & * & 0 \\
0 & 0 & *
\end{pmatrix}$

$\begin{pmatrix}
* & * & 0 \\
* & * & 0 \\
0 & 0 & *
\end{pmatrix}$
&
$\begin{pmatrix}
* & 0 & * \\
* & 0 & * \\
0 & * & 0
\end{pmatrix}$

$\begin{pmatrix}
* & * & 0 \\
* & * & 0 \\
0 & 0 & *
\end{pmatrix}$
&
$\begin{pmatrix}
0 & * & 0 \\
0 & * & 0 \\
* & 0 & 0
\end{pmatrix}$

$\begin{pmatrix}
* & * & 0 \\
* & * & 0 \\
0 & 0 & *
\end{pmatrix}$
\\ \hline

\end{tabular}
    \caption{$Y^{ij}$ and  ${C_{\mathrm{W}}}^{ij}$ matrices for $M=4$.}
    \label{tab:Y-C-M=4-3}
\end{table}

\newpage

\begin{table}[H]
    \centering
    \begin{tabular}{|c|c|c|c|} \hline
        Flavor & Higgs\ $[g^0]$ & Higgs\ $[g^1]$ &Higgs\ $[g^2]$ \\ 
        (Left,Right) & $Y^{ij}$,\hspace{1cm} ${C_{\mathrm{W}}}^{ij}$ & $Y^{ij}$,\hspace{1cm} ${C_{\mathrm{W}}}^{ij}$ & $Y^{ij}$,\hspace{1cm} ${C_{\mathrm{W}}}^{ij}$\\\hline

(\text{vi},\text{ii}) & $\begin{pmatrix}
0 & 0 & * \\
0 & 0 & * \\
0 & 0 & 0
\end{pmatrix}$

$\begin{pmatrix}
* & * & 0 \\
* & * & 0 \\
0 & 0 & *
\end{pmatrix}$
&
$\begin{pmatrix}
* & * & 0 \\
* & * & 0 \\
0 & 0 & *
\end{pmatrix}$

$\begin{pmatrix}
* & * & 0 \\
* & * & 0 \\
0 & 0 & *
\end{pmatrix}$
&
$\begin{pmatrix}
0 & 0 & * \\
0 & 0 & * \\
* & * & 0
\end{pmatrix}$

$\begin{pmatrix}
* & * & 0 \\
* & * & 0 \\
0 & 0 & *
\end{pmatrix}$
\\ \hline

(\text{vi},\text{iii})  & $\begin{pmatrix}
0 & 0 & 0 \\
0 & 0 & 0 \\
0 & 0 & *
\end{pmatrix}$

$\begin{pmatrix}
* & * & 0 \\
* & * & 0 \\
0 & 0 & *
\end{pmatrix}$
&
$\begin{pmatrix}
* & * & * \\
* & * & * \\
0 & 0 & 0
\end{pmatrix}$

$\begin{pmatrix}
* & * & 0 \\
* & * & 0 \\
0 & 0 & *
\end{pmatrix}$
&
$\begin{pmatrix}
0 & 0 & 0 \\
0 & 0 & 0 \\
* & * & 0
\end{pmatrix}$

$\begin{pmatrix}
* & * & 0 \\
* & * & 0 \\
0 & 0 & *
\end{pmatrix}$
\\ \hline

(\text{vi},\text{iv})   & $\begin{pmatrix}
0 & * & * \\
0 & * & * \\
0 & 0 & 0
\end{pmatrix}$

$\begin{pmatrix}
* & * & 0 \\
* & * & 0 \\
0 & 0 & *
\end{pmatrix}$
&
$\begin{pmatrix}
* & 0 & 0 \\
* & 0 & 0 \\
0 & * & *
\end{pmatrix}$

$\begin{pmatrix}
* & * & 0 \\
* & * & 0 \\
0 & 0 & *
\end{pmatrix}$
&
$\begin{pmatrix}
0 & * & * \\
0 & * & * \\
* & 0 & 0
\end{pmatrix}$

$\begin{pmatrix}
* & * & 0 \\
* & * & 0 \\
0 & 0 & *
\end{pmatrix}$
\\ \hline

(\text{vi},\text{v}) 
&
$\begin{pmatrix}
0 & 0 & 0 \\
0 & 0 & 0 \\
0 & * & *
\end{pmatrix}$

$\begin{pmatrix}
* & * & 0 \\
* & * & 0 \\
0 & 0 & *
\end{pmatrix}$
&
$\begin{pmatrix}
* & * & * \\
* & * & * \\
0 & 0 & 0
\end{pmatrix}$

$\begin{pmatrix}
* & * & 0 \\
* & * & 0 \\
0 & 0 & *
\end{pmatrix}$
&
$\begin{pmatrix}
0 & 0 & 0 \\
0 & 0 & 0 \\
* & 0 & 0
\end{pmatrix}$

$\begin{pmatrix}
* & * & 0 \\
* & * & 0 \\
0 & 0 & *
\end{pmatrix}$
\\ \hline

(\text{vi},\text{vi})  & $\begin{pmatrix}
* & * & 0 \\
* & * & 0 \\
0 & 0 & *
\end{pmatrix}$

$\begin{pmatrix}
* & * & 0 \\
* & * & 0 \\
0 & 0 & *
\end{pmatrix}$
&
$\begin{pmatrix}
0 & 0 & * \\
0 & 0 & * \\
* & * & 0
\end{pmatrix}$

$\begin{pmatrix}
* & * & 0 \\
* & * & 0 \\
0 & 0 & *
\end{pmatrix}$
&
$\begin{pmatrix}
* & * & 0 \\
* & * & 0 \\
0 & 0 & 0
\end{pmatrix}$

$\begin{pmatrix}
* & * & 0 \\
* & * & 0 \\
0 & 0 & *
\end{pmatrix}$
\\ \hline

(\text{vi},\text{vii})  
&
$\begin{pmatrix}
* & 0 & 0 \\
* & 0 & 0 \\
0 & * & *
\end{pmatrix}$

$\begin{pmatrix}
* & * & 0 \\
* & * & 0 \\
0 & 0 & *
\end{pmatrix}$
&
$\begin{pmatrix}
0 & * & * \\
0 & * & * \\
* & 0 & 0
\end{pmatrix}$

$\begin{pmatrix}
* & * & 0 \\
* & * & 0 \\
0 & 0 & *
\end{pmatrix}$
&
$\begin{pmatrix}
* & 0 & 0 \\
* & 0 & 0 \\
0 & 0 & 0
\end{pmatrix}$

$\begin{pmatrix}
* & * & 0 \\
* & * & 0 \\
0 & 0 & *
\end{pmatrix}$
\\ \hline

(\text{vii},\text{i})  & $\begin{pmatrix}
0 & * & 0 \\
0 & 0 & * \\
0 & 0 & *
\end{pmatrix}$

$\begin{pmatrix}
* & 0 & 0 \\
0 & * & * \\
0 & * & *
\end{pmatrix}$
&
$\begin{pmatrix}
* & 0 & * \\
0 & * & 0 \\
0 & * & 0
\end{pmatrix}$

$\begin{pmatrix}
* & 0 & 0 \\
0 & * & * \\
0 & * & *
\end{pmatrix}$
&
$\begin{pmatrix}
0 & * & 0 \\
* & 0 & 0 \\
* & 0 & 0
\end{pmatrix}$

$\begin{pmatrix}
* & 0 & 0 \\
0 & * & * \\
0 & * & *
\end{pmatrix}$
\\ \hline

(\text{vii},\text{ii}) &  $\begin{pmatrix}
0 & 0 & * \\
0 & 0 & 0 \\
0 & 0 & 0
\end{pmatrix}$

$\begin{pmatrix}
* & 0 & 0 \\
0 & * & * \\
0 & * & *
\end{pmatrix}$
&
$\begin{pmatrix}
* & * & 0 \\
0 & 0 & * \\
0 & 0 & *
\end{pmatrix}$

$\begin{pmatrix}
* & 0 & 0 \\
0 & * & * \\
0 & * & *
\end{pmatrix}$
&
$\begin{pmatrix}
0 & 0 & * \\
* & * & 0 \\
* & * & 0
\end{pmatrix}$

$\begin{pmatrix}
* & 0 & 0 \\
0 & * & * \\
0 & * & *
\end{pmatrix}$
\\ \hline

(\text{vii},\text{iii})   & $\begin{pmatrix}
0 & 0 & 0 \\
0 & 0 & * \\
0 & 0 & *
\end{pmatrix}$

$\begin{pmatrix}
* & 0 & 0 \\
0 & * & * \\
0 & * & *
\end{pmatrix}$
&
$\begin{pmatrix}
* & * & * \\
0 & 0 & 0 \\
0 & 0 & 0
\end{pmatrix}$

$\begin{pmatrix}
* & 0 & 0 \\
0 & * & * \\
0 & * & *
\end{pmatrix}$
&
$\begin{pmatrix}
0 & 0 & 0 \\
* & * & 0 \\
* & * & 0
\end{pmatrix}$

$\begin{pmatrix}
* & 0 & 0 \\
0 & * & * \\
0 & * & *
\end{pmatrix}$
\\ \hline

(\text{vii},\text{iv})   & $\begin{pmatrix}
0 & * & * \\
0 & 0 & 0 \\
0 & 0 & 0
\end{pmatrix}$

$\begin{pmatrix}
* & 0 & 0 \\
0 & * & * \\
0 & * & *
\end{pmatrix}$
&
$\begin{pmatrix}
* & 0 & 0 \\
0 & * & * \\
0 & * & *
\end{pmatrix}$

$\begin{pmatrix}
* & 0 & 0 \\
0 & * & * \\
0 & * & *
\end{pmatrix}$
&
$\begin{pmatrix}
0 & * & * \\
* & 0 & 0 \\
* & 0 & 0
\end{pmatrix}$

$\begin{pmatrix}
* & 0 & 0 \\
0 & * & * \\
0 & * & *
\end{pmatrix}$
\\ \hline

(\text{vii},\text{v}) 
& 
$\begin{pmatrix}
0 & 0 & 0 \\
0 & * & * \\
0 & * & *
\end{pmatrix}$

$\begin{pmatrix}
* & 0 & 0 \\
0 & * & * \\
0 & * & *
\end{pmatrix}$
&
$\begin{pmatrix}
* & * & * \\
0 & 0 & 0 \\
0 & 0 & 0
\end{pmatrix}$

$\begin{pmatrix}
* & 0 & 0 \\
0 & * & * \\
0 & * & *
\end{pmatrix}$
&
$\begin{pmatrix}
0 & 0 & 0 \\
* & 0 & 0 \\
* & 0 & 0
\end{pmatrix}$

$\begin{pmatrix}
* & 0 & 0 \\
0 & * & * \\
0 & * & *
\end{pmatrix}$
\\ \hline

(\text{vii},\text{vi})  
&
$\begin{pmatrix}
* & * & 0 \\
0 & 0 & * \\
0 & 0 & *
\end{pmatrix}$

$\begin{pmatrix}
* & 0 & 0 \\
0 & * & * \\
0 & * & *
\end{pmatrix}$
&
$\begin{pmatrix}
0 & 0 & * \\
* & * & 0 \\
* & * & 0
\end{pmatrix}$

$\begin{pmatrix}
* & 0 & 0 \\
0 & * & * \\
0 & * & *
\end{pmatrix}$
&
$\begin{pmatrix}
* & * & 0 \\
0 & 0 & 0 \\
0 & 0 & 0
\end{pmatrix}$

$\begin{pmatrix}
* & 0 & 0 \\
0 & * & * \\
0 & * & *
\end{pmatrix}$
\\ \hline
 \end{tabular}
    \caption{$Y^{ij}$ and ${C_{\mathrm{W}}}^{ij}$ matrices for $M=4$.}
    \label{tab:Y-C-M=4-4}
\end{table}

\newpage

\begin{table}[H]
    \centering
    \begin{tabular}{|c|c|c|c|} \hline
        Flavor & Higgs\ $[g^0]$ & Higgs\ $[g^1]$ &Higgs\ $[g^2]$ \\ 
        (Left,Right) & $Y^{ij}$,\hspace{1cm} ${C_{\mathrm{W}}}^{ij}$ & $Y^{ij}$,\hspace{1cm} ${C_{\mathrm{W}}}^{ij}$ & $Y^{ij}$,\hspace{1cm} ${C_{\mathrm{W}}}^{ij}$\\\hline
        
(\text{vii},\text{vii})  
&
$\begin{pmatrix}
* & 0 & 0 \\
0 & * & * \\
0 & * & *
\end{pmatrix}$

$\begin{pmatrix}
* & 0 & 0 \\
0 & * & * \\
0 & * & *
\end{pmatrix}$
&
$\begin{pmatrix}
0 & * & * \\
* & 0 & 0 \\
* & 0 & 0
\end{pmatrix}$

$\begin{pmatrix}
* & 0 & 0 \\
0 & * & * \\
0 & * & *
\end{pmatrix}$
&
$\begin{pmatrix}
* & 0 & 0 \\
0 & 0 & 0 \\
0 & 0 & 0
\end{pmatrix}$

$\begin{pmatrix}
* & 0 & 0 \\
0 & * & * \\
0 & * & *
\end{pmatrix}$
\\ \hline

    \end{tabular}
    \caption{$Y^{ij}$ and ${C_{\mathrm{W}}}^{ij}$ matrices for $M=4$.}
    \label{tab:Y-C-M=4-5}
\end{table}

\subsubsection{$M=5$}
\label{sec:Y-C-M=5}

We study the case with $M=5$.
We use the same flavor assignments for $e_i$ as ones for $L_i$ in Table~\ref{tab:Flavor-M=5}.
The coupling selection rules discussed in section \ref{sec:Z2-gauging} allow the following couplings:
\begin{align}
    \phi_0 \phi_0 \phi_0, \qquad \phi_0 \phi_1 \phi_1, \qquad \phi_0 \phi_2 \phi_2 ,\qquad \phi_1 \phi_1 \phi_2, \qquad \phi_1 \phi_2 \phi_2,
\end{align}
where $\phi_k$ corresponds to the class $[g^k]$.
By use of these selection rules, we can examine the possible textures of $Y^{ij}$.
Tables~\ref{tab:Y-C-M=5-1},\,\ref{tab:Y-C-M=5-2},\,\ref{tab:Y-C-M=5-3},\,\ref{tab:Y-C-M=5-4},\,\ref{tab:Y-C-M=5-5} show possible combinations of $Y^{ij}$ and ${C_{\mathrm{W}}}^{ij}$.
In the Standard Model, the Higgs mode corresponds to a single class.
However, $H_u$ and $H_d$ can correspond to different classes in supersymmetric standard models and two Higgs doublet models.

\newpage

\begin{table}[H]
    \centering
    \begin{tabular}{|c|c|c|c|} \hline
        Flavor & Higgs\ $[g^0]$ & Higgs\ $[g^1]$ &Higgs\ $[g^2]$ \\ 
        (Left,Right) & $Y^{ij}$,\hspace{1cm} ${C_{\mathrm{W}}}^{ij}$ & $Y^{ij}$,\hspace{1cm} ${C_{\mathrm{W}}}^{ij}$ & $Y^{ij}$,\hspace{1cm} ${C_{\mathrm{W}}}^{ij}$\\\hline
        
(\text{i},\text{i})  
&
$\begin{pmatrix}
* & 0 & 0 \\
0 & * & 0 \\
0 & 0 & *
\end{pmatrix}$

$\begin{pmatrix}
* & 0 & 0 \\
0 & * & 0 \\
0 & 0 & *
\end{pmatrix}$
&
$\begin{pmatrix}
0 & * & 0 \\
* & 0 & * \\
0 & * & *
\end{pmatrix}$

$\begin{pmatrix}
* & 0 & * \\
0 & * & * \\
* & * & *
\end{pmatrix}$
&
$\begin{pmatrix}
0 & 0 & * \\
0 & * & * \\
* & * & 0
\end{pmatrix}$

$\begin{pmatrix}
* & * & 0 \\
* & * & * \\
0 & * & *
\end{pmatrix}$
\\ \hline

(\text{i},\text{ii})  
&
$\begin{pmatrix}
* & * & 0 \\
0 & 0 & * \\
0 & 0 & 0
\end{pmatrix}$

$\begin{pmatrix}
* & 0 & 0 \\
0 & * & 0 \\
0 & 0 & *
\end{pmatrix}$
&
$\begin{pmatrix}
0 & 0 & * \\
* & * & 0 \\
0 & 0 & *
\end{pmatrix}$

$\begin{pmatrix}
* & 0 & * \\
0 & * & * \\
* & * & *
\end{pmatrix}$
&
$\begin{pmatrix}
0 & 0 & 0 \\
0 & 0 & * \\
* & * & *
\end{pmatrix}$

$\begin{pmatrix}
* & * & 0 \\
* & * & * \\
0 & * & *
\end{pmatrix}$
\\ \hline

(\text{i},\text{iii})  
&
$\begin{pmatrix}
* & * & 0 \\
0 & 0 & 0 \\
0 & 0 & *
\end{pmatrix}$

$\begin{pmatrix}
* & 0 & 0 \\
0 & * & 0 \\
0 & 0 & *
\end{pmatrix}$
&
$\begin{pmatrix}
0 & 0 & 0 \\
* & * & * \\
0 & 0 & *
\end{pmatrix}$

$\begin{pmatrix}
* & 0 & * \\
0 & * & * \\
* & * & *
\end{pmatrix}$
&
$\begin{pmatrix}
0 & 0 & * \\
0 & 0 & * \\
* & * & 0
\end{pmatrix}$

$\begin{pmatrix}
* & * & 0 \\
* & * & * \\
0 & * & *
\end{pmatrix}$
\\ \hline

(\text{i},\text{iv})  
&
$\begin{pmatrix}
* & 0 & 0 \\
0 & * & * \\
0 & 0 & 0
\end{pmatrix}$

$\begin{pmatrix}
* & 0 & 0 \\
0 & * & 0 \\
0 & 0 & *
\end{pmatrix}$
&
$\begin{pmatrix}
0 & * & * \\
* & 0 & 0 \\
0 & * & *
\end{pmatrix}$

$\begin{pmatrix}
* & 0 & * \\
0 & * & * \\
* & * & *
\end{pmatrix}$
&
$\begin{pmatrix}
0 & 0 & 0 \\
0 & * & * \\
* & * & *
\end{pmatrix}$

$\begin{pmatrix}
* & * & 0 \\
* & * & * \\
0 & * & *
\end{pmatrix}$
\\ \hline

(\text{i},\text{v})  
&
$\begin{pmatrix}
* & 0 & 0 \\
0 & 0 & 0 \\
0 & * & *
\end{pmatrix}$

$\begin{pmatrix}
* & 0 & 0 \\
0 & * & 0 \\
0 & 0 & *
\end{pmatrix}$
&
$\begin{pmatrix}
0 & 0 & 0 \\
* & * & * \\
0 & * & *
\end{pmatrix}$

$\begin{pmatrix}
* & 0 & * \\
0 & * & * \\
* & * & *
\end{pmatrix}$
&
$\begin{pmatrix}
0 & * & * \\
0 & * & * \\
* & 0 & 0
\end{pmatrix}$

$\begin{pmatrix}
* & * & 0 \\
* & * & * \\
0 & * & *
\end{pmatrix}$
\\ \hline

(\text{i},\text{vi})  
&
$\begin{pmatrix}
0 & 0 & 0 \\
* & * & 0 \\
0 & 0 & *
\end{pmatrix}$

$\begin{pmatrix}
* & 0 & 0 \\
0 & * & 0 \\
0 & 0 & *
\end{pmatrix}$
&
$\begin{pmatrix}
* & * & 0 \\
0 & 0 & * \\
* & * & *
\end{pmatrix}$

$\begin{pmatrix}
* & 0 & * \\
0 & * & * \\
* & * & *
\end{pmatrix}$
&
$\begin{pmatrix}
0 & 0 & * \\
* & * & * \\
* & * & 0
\end{pmatrix}$

$\begin{pmatrix}
* & * & 0 \\
* & * & * \\
0 & * & *
\end{pmatrix}$
\\ \hline

(\text{i},\text{vii})  
&
$\begin{pmatrix}
0 & 0 & 0 \\
* & 0 & 0 \\
0 & * & *
\end{pmatrix}$

$\begin{pmatrix}
* & 0 & 0 \\
0 & * & 0 \\
0 & 0 & *
\end{pmatrix}$
&
$\begin{pmatrix}
* & 0 & 0 \\
0 & * & * \\
* & * & *
\end{pmatrix}$

$\begin{pmatrix}
* & 0 & * \\
0 & * & * \\
* & * & *
\end{pmatrix}$
&
$\begin{pmatrix}
0 & * & * \\
* & * & * \\
* & 0 & 0
\end{pmatrix}$

$\begin{pmatrix}
* & * & 0 \\
* & * & * \\
0 & * & *
\end{pmatrix}$
\\ \hline

(\text{ii},\text{i})  
&
$\begin{pmatrix}
* & 0 & 0 \\
* & 0 & 0 \\
0 & * & 0
\end{pmatrix}$

$\begin{pmatrix}
* & * & 0 \\
* & * & 0 \\
0 & 0 & *
\end{pmatrix}$
&
$\begin{pmatrix}
0 & * & 0 \\
0 & * & 0 \\
* & 0 & *
\end{pmatrix}$

$\begin{pmatrix}
* & * & 0 \\
* & * & 0 \\
0 & 0 & *
\end{pmatrix}$
&
$\begin{pmatrix}
0 & 0 & * \\
0 & 0 & * \\
0 & * & *
\end{pmatrix}$

$\begin{pmatrix}
* & * & * \\
* & * & * \\
* & * & *
\end{pmatrix}$
\\ \hline

(\text{ii},\text{ii})  
&
$\begin{pmatrix}
* & * & 0 \\
* & * & 0 \\
0 & 0 & *
\end{pmatrix}$

$\begin{pmatrix}
* & * & 0 \\
* & * & 0 \\
0 & 0 & *
\end{pmatrix}$
&
$\begin{pmatrix}
0 & 0 & * \\
0 & 0 & * \\
* & * & 0
\end{pmatrix}$

$\begin{pmatrix}
* & * & 0 \\
* & * & 0 \\
0 & 0 & *
\end{pmatrix}$
&
$\begin{pmatrix}
0 & 0 & 0 \\
0 & 0 & 0 \\
0 & 0 & *
\end{pmatrix}$

$\begin{pmatrix}
* & * & * \\
* & * & * \\
* & * & *
\end{pmatrix}$
\\ \hline

(\text{ii},\text{iii})  
&
$\begin{pmatrix}
* & * & 0 \\
* & * & 0 \\
0 & 0 & 0
\end{pmatrix}$

$\begin{pmatrix}
* & * & 0 \\
* & * & 0 \\
0 & 0 & *
\end{pmatrix}$
&
$\begin{pmatrix}
0 & 0 & 0 \\
0 & 0 & 0 \\
* & * & *
\end{pmatrix}$

$\begin{pmatrix}
* & * & 0 \\
* & * & 0 \\
0 & 0 & *
\end{pmatrix}$
&
$\begin{pmatrix}
0 & 0 & *  \\
0 & 0 & * \\
0 & 0 & *
\end{pmatrix}$

$\begin{pmatrix}
* & * & * \\
* & * & * \\
* & * & *
\end{pmatrix}$
\\ \hline

(\text{ii},\text{iv})  
&
$\begin{pmatrix}
* & 0 & 0 \\
* & 0 & 0 \\
0 & * & *
\end{pmatrix}$

$\begin{pmatrix}
* & * & 0 \\
* & * & 0 \\
0 & 0 & *
\end{pmatrix}$
&
$\begin{pmatrix}
0 & * & * \\
0 & * & * \\
* & 0 & 0
\end{pmatrix}$

$\begin{pmatrix}
* & * & 0 \\
* & * & 0 \\
0 & 0 & *
\end{pmatrix}$
&
$\begin{pmatrix}
0 & 0 & 0 \\
0 & 0 & 0 \\
0 & * & *
\end{pmatrix}$

$\begin{pmatrix}
* & * & * \\
* & * & * \\
* & * & *
\end{pmatrix}$
\\ \hline

(\text{ii},\text{v})  
&
$\begin{pmatrix}
* & 0 & 0 \\
* & 0 & 0 \\
0 & 0 & 0
\end{pmatrix}$

$\begin{pmatrix}
* & * & 0 \\
* & * & 0 \\
0 & 0 & *
\end{pmatrix}$
&
$\begin{pmatrix}
0 & 0 & 0 \\
0 & 0 & 0 \\
* & * & *
\end{pmatrix}$

$\begin{pmatrix}
* & * & 0 \\
* & * & 0 \\
0 & 0 & *
\end{pmatrix}$
&
$\begin{pmatrix}
0 & * & * \\
0 & * & * \\
0 & * & *
\end{pmatrix}$

$\begin{pmatrix}
* & * & * \\
* & * & * \\
* & * & *
\end{pmatrix}$
\\ \hline

    \end{tabular}
    \caption{$Y^{ij}$ and ${C_{\mathrm{W}}}^{ij}$ matrices for $M=5$.}
    \label{tab:Y-C-M=5-1}
\end{table}

\newpage

\begin{table}[H]
    \centering
    \begin{tabular}{|c|c|c|c|} \hline
        Flavor & Higgs\ $[g^0]$ & Higgs\ $[g^1]$ &Higgs\ $[g^2]$ \\ 
        (Left,Right) & $Y^{ij}$,\hspace{1cm} ${C_{\mathrm{W}}}^{ij}$ & $Y^{ij}$,\hspace{1cm} ${C_{\mathrm{W}}}^{ij}$ & $Y^{ij}$,\hspace{1cm} ${C_{\mathrm{W}}}^{ij}$\\\hline

(\text{ii},\text{vi})  
&
$\begin{pmatrix}
0 & 0 & 0 \\
0 & 0 & 0 \\
* & * & 0
\end{pmatrix}$

$\begin{pmatrix}
* & * & 0 \\
* & * & 0 \\
0 & 0 & *
\end{pmatrix}$
&
$\begin{pmatrix}
* & * & 0 \\
* & * & 0 \\
0 & 0 & *
\end{pmatrix}$

$\begin{pmatrix}
* & * & 0 \\
* & * & 0 \\
0 & 0 & *
\end{pmatrix}$
&
$\begin{pmatrix}
0 & 0 & * \\
0 & 0 & * \\
* & * & *
\end{pmatrix}$

$\begin{pmatrix}
* & * & * \\
* & * & * \\
* & * & *
\end{pmatrix}$
\\ \hline

(\text{ii},\text{vii})  
&
$\begin{pmatrix}
0 & 0 & 0 \\
0 & 0 & 0 \\
* & 0 & 0
\end{pmatrix}$

$\begin{pmatrix}
* & * & 0 \\
* & * & 0 \\
0 & 0 & *
\end{pmatrix}$
&
$\begin{pmatrix}
* & 0 & 0 \\
* & 0 & 0 \\
0 & * & *
\end{pmatrix}$

$\begin{pmatrix}
* & * & 0 \\
* & * & 0 \\
0 & 0 & *
\end{pmatrix}$
&
$\begin{pmatrix}
0 & * & * \\
0 & * & * \\
* & * & *
\end{pmatrix}$

$\begin{pmatrix}
* & * & * \\
* & * & * \\
* & * & *
\end{pmatrix}$
\\ \hline

(\text{iii},\text{i})  
&
$\begin{pmatrix}
* & 0 & 0 \\
* & 0 & 0 \\
0 & 0 & *
\end{pmatrix}$

$\begin{pmatrix}
* & * & 0 \\
* & * & 0 \\
0 & 0 & *
\end{pmatrix}$
&
$\begin{pmatrix}
0 & * & 0 \\
0 & * & 0 \\
0 & * & *
\end{pmatrix}$

$\begin{pmatrix}
* & * & * \\
* & * & * \\
* & * & *
\end{pmatrix}$
&
$\begin{pmatrix}
0 & 0 & * \\
0 & 0 & * \\
* & * & 0
\end{pmatrix}$

$\begin{pmatrix}
* & * & 0 \\
* & * & 0 \\
0 & 0 & *
\end{pmatrix}$
\\ \hline

(\text{iii},\text{ii})  
&
$\begin{pmatrix}
* & * & 0 \\
* & * & 0 \\
0 & 0 & 0
\end{pmatrix}$

$\begin{pmatrix}
* & * & 0 \\
* & * & 0 \\
0 & 0 & *
\end{pmatrix}$
&
$\begin{pmatrix}
0 & 0 & * \\
0 & 0 & * \\
0 & 0 & *
\end{pmatrix}$

$\begin{pmatrix}
* & * & * \\
* & * & * \\
* & * & *
\end{pmatrix}$
&
$\begin{pmatrix}
0 & 0 & 0 \\
0 & 0 & 0 \\
* & * & *
\end{pmatrix}$

$\begin{pmatrix}
* & * & 0 \\
* & * & 0 \\
0 & 0 & *
\end{pmatrix}$
\\ \hline

(\text{iii},\text{iii})  
&
$\begin{pmatrix}
* & * & 0 \\
* & * & 0 \\
0 & 0 & *
\end{pmatrix}$

$\begin{pmatrix}
* & * & 0 \\
* & * & 0 \\
0 & 0 & *
\end{pmatrix}$
&
$\begin{pmatrix}
0 & 0 & 0 \\
0 & 0 & 0 \\
0 & 0 & *
\end{pmatrix}$

$\begin{pmatrix}
* & * & * \\
* & * & * \\
* & * & *
\end{pmatrix}$
&
$\begin{pmatrix}
0 & 0 & * \\
0 & 0 & * \\
* & * & 0
\end{pmatrix}$

$\begin{pmatrix}
* & * & 0 \\
* & * & 0 \\
0 & 0 & *
\end{pmatrix}$
\\ \hline

(\text{iii},\text{iv})  
&
$\begin{pmatrix}
* & 0 & 0 \\
* & 0 & 0 \\
0 & 0 & 0
\end{pmatrix}$

$\begin{pmatrix}
* & * & 0 \\
* & * & 0 \\
0 & 0 & *
\end{pmatrix}$
&
$\begin{pmatrix}
0 & * & * \\
0 & * & * \\
0 & * & *
\end{pmatrix}$

$\begin{pmatrix}
* & * & * \\
* & * & * \\
* & * & *
\end{pmatrix}$
&
$\begin{pmatrix}
0 & 0 & 0 \\
0 & 0 & 0 \\
* & * & *
\end{pmatrix}$

$\begin{pmatrix}
* & * & 0 \\
* & * & 0 \\
0 & 0 & *
\end{pmatrix}$
\\ \hline

(\text{iii},\text{v})  
&
$\begin{pmatrix}
* & 0 & 0 \\
* & 0 & 0 \\
0 & * & *
\end{pmatrix}$

$\begin{pmatrix}
* & * & 0 \\
* & * & 0 \\
0 & 0 & *
\end{pmatrix}$
&
$\begin{pmatrix}
0 & 0 & 0 \\
0 & 0 & 0 \\
0 & * & *
\end{pmatrix}$

$\begin{pmatrix}
* & * & * \\
* & * & * \\
* & * & *
\end{pmatrix}$
&
$\begin{pmatrix}
0 & * & * \\
0 & * & * \\
* & 0 & 0
\end{pmatrix}$

$\begin{pmatrix}
* & * & 0 \\
* & * & 0 \\
0 & 0 & *
\end{pmatrix}$
\\ \hline

(\text{iii},\text{vi})  
&
$\begin{pmatrix}
0 & 0 & 0 \\
0 & 0 & 0 \\
0 & 0 & *
\end{pmatrix}$

$\begin{pmatrix}
* & * & 0 \\
* & * & 0 \\
0 & 0 & *
\end{pmatrix}$
&
$\begin{pmatrix}
* & * & 0 \\
* & * & 0 \\
* & * & *
\end{pmatrix}$

$\begin{pmatrix}
* & * & * \\
* & * & * \\
* & * & *
\end{pmatrix}$
&
$\begin{pmatrix}
0 & 0 & * \\
0 & 0 & * \\
* & * & 0
\end{pmatrix}$

$\begin{pmatrix}
* & * & 0 \\
* & * & 0 \\
0 & 0 & *
\end{pmatrix}$
\\ \hline

(\text{iii},\text{vii})  
&
$\begin{pmatrix}
0 & 0 & 0 \\
0 & 0 & 0 \\
0 & * & *
\end{pmatrix}$

$\begin{pmatrix}
* & * & 0 \\
* & * & 0 \\
0 & 0 & *
\end{pmatrix}$
&
$\begin{pmatrix}
* & 0 & 0 \\
* & 0 & 0 \\
* & * & *
\end{pmatrix}$

$\begin{pmatrix}
* & * & * \\
* & * & * \\
* & * & *
\end{pmatrix}$
&
$\begin{pmatrix}
0 & * & * \\
0 & * & * \\
* & 0 & 0
\end{pmatrix}$

$\begin{pmatrix}
* & * & 0 \\
* & * & 0 \\
0 & 0 & *
\end{pmatrix}$
\\ \hline

(\text{iv},\text{i})  
&
$\begin{pmatrix}
* & 0 & 0 \\
0 & * & 0 \\
0 & * & 0
\end{pmatrix}$

$\begin{pmatrix}
* & 0 & 0 \\
0 & * & * \\
0 & * & *
\end{pmatrix}$
&
$\begin{pmatrix}
0 & * & 0 \\
* & 0 & * \\
* & 0 & *
\end{pmatrix}$

$\begin{pmatrix}
* & 0 & 0 \\
0 & * & * \\
0 & * & *
\end{pmatrix}$
&
$\begin{pmatrix}
0 & 0 & * \\
0 & * & * \\
0 & * & *
\end{pmatrix}$

$\begin{pmatrix}
* & * & * \\
* & * & * \\
* & * & *
\end{pmatrix}$
\\ \hline

(\text{iv},\text{ii})  
&
$\begin{pmatrix}
* & * & 0 \\
0 & 0 & * \\
0 & 0 & *
\end{pmatrix}$

$\begin{pmatrix}
* & 0 & 0 \\
0 & * & * \\
0 & * & *
\end{pmatrix}$
&
$\begin{pmatrix}
0 & 0 & * \\
* & * & 0 \\
* & * & 0
\end{pmatrix}$

$\begin{pmatrix}
* & 0 & 0 \\
0 & * & * \\
0 & * & *
\end{pmatrix}$
&
$\begin{pmatrix}
0 & 0 & 0 \\
0 & 0 & * \\
0 & 0 & *
\end{pmatrix}$

$\begin{pmatrix}
* & * & * \\
* & * & * \\
* & * & *
\end{pmatrix}$
\\ \hline

(\text{iv},\text{iii})  
&
$\begin{pmatrix}
* & * & 0 \\
0 & 0 & 0 \\
0 & 0 & 0
\end{pmatrix}$

$\begin{pmatrix}
* & 0 & 0 \\
0 & * & * \\
0 & * & *
\end{pmatrix}$
&
$\begin{pmatrix}
0 & 0 & 0 \\
* & * & * \\
* & * & *
\end{pmatrix}$

$\begin{pmatrix}
* & 0 & 0 \\
0 & * & * \\
0 & * & *
\end{pmatrix}$
&
$\begin{pmatrix}
0 & 0 & * \\
0 & 0 & * \\
0 & 0 & *
\end{pmatrix}$

$\begin{pmatrix}
* & * & * \\
* & * & * \\
* & * & *
\end{pmatrix}$
\\ \hline

\end{tabular}
    \caption{$Y^{ij}$ and ${C_{\mathrm{W}}}^{ij}$ matrices for $M=5$.}
    \label{tab:Y-C-M=5-2}
\end{table}

\begin{table}[H]
    \centering
    \begin{tabular}{|c|c|c|c|} \hline
        Flavor & Higgs\ $[g^0]$ & Higgs\ $[g^1]$ &Higgs\ $[g^2]$ \\ 
        (Left,Right) & $Y^{ij}$,\hspace{1cm} ${C_{\mathrm{W}}}^{ij}$ & $Y^{ij}$,\hspace{1cm} ${C_{\mathrm{W}}}^{ij}$ & $Y^{ij}$,\hspace{1cm} ${C_{\mathrm{W}}}^{ij}$\\\hline

(\text{iv},\text{iv})  
&
$\begin{pmatrix}
* & 0 & 0 \\
0 & * & * \\
0 & * & *
\end{pmatrix}$

$\begin{pmatrix}
* & 0 & 0 \\
0 & * & * \\
0 & * & *
\end{pmatrix}$
&
$\begin{pmatrix}
0 & * & * \\
* & 0 & 0 \\
* & 0 & 0
\end{pmatrix}$

$\begin{pmatrix}
* & 0 & 0 \\
0 & * & * \\
0 & * & *
\end{pmatrix}$
&
$\begin{pmatrix}
0 & 0 & 0\\
0 & * & * \\
0 & * & *
\end{pmatrix}$

$\begin{pmatrix}
* & * & * \\
* & * & * \\
* & * & *
\end{pmatrix}$
\\ \hline

(\text{iv},\text{v})  
&
$\begin{pmatrix}
* & 0 & 0 \\
0 & 0 & 0 \\
0 & 0 & 0
\end{pmatrix}$

$\begin{pmatrix}
* & 0 & 0 \\
0 & * & * \\
0 & * & *
\end{pmatrix}$
&
$\begin{pmatrix}
0 & 0 & 0 \\
* & * & * \\
* & * & *
\end{pmatrix}$

$\begin{pmatrix}
* & 0 & 0 \\
0 & * & * \\
0 & * & *
\end{pmatrix}$
&
$\begin{pmatrix}
0 & * & *\\
0 & * & * \\
0 & * & *
\end{pmatrix}$

$\begin{pmatrix}
* & * & * \\
* & * & * \\
* & * & *
\end{pmatrix}$
\\ \hline

(\text{iv},\text{vi})  
&
$\begin{pmatrix}
0 & 0 & 0 \\
* & * & 0 \\
* & * & 0
\end{pmatrix}$

$\begin{pmatrix}
* & 0 & 0 \\
0 & * & * \\
0 & * & *
\end{pmatrix}$
&
$\begin{pmatrix}
* & * & 0 \\
0 & 0 & * \\
0 & 0 & *
\end{pmatrix}$

$\begin{pmatrix}
* & 0 & 0 \\
0 & * & * \\
0 & * & *
\end{pmatrix}$
&
$\begin{pmatrix}
0 & 0 & *\\
* & * & * \\
* & * & *
\end{pmatrix}$

$\begin{pmatrix}
* & * & * \\
* & * & * \\
* & * & *
\end{pmatrix}$
\\ \hline

(\text{iv},\text{vii})  
&
$\begin{pmatrix}
0 & 0 & 0 \\
* & 0 & 0 \\
* & 0 & 0
\end{pmatrix}$

$\begin{pmatrix}
* & 0 & 0 \\
0 & * & * \\
0 & * & *
\end{pmatrix}$
&
$\begin{pmatrix}
* & 0 & 0 \\
0 & * & * \\
0 & * & *
\end{pmatrix}$

$\begin{pmatrix}
* & 0 & 0 \\
0 & * & * \\
0 & * & *
\end{pmatrix}$
&
$\begin{pmatrix}
0 & * & *\\
* & * & * \\
* & * & *
\end{pmatrix}$

$\begin{pmatrix}
* & * & * \\
* & * & * \\
* & * & *
\end{pmatrix}$
\\ \hline

(\text{v},\text{i})  
&
$\begin{pmatrix}
* & 0 & 0 \\
0 & 0 & * \\
0 & 0 & *
\end{pmatrix}$

$\begin{pmatrix}
* & 0 & 0 \\
0 & * & * \\
0 & * & *
\end{pmatrix}$
&
$\begin{pmatrix}
0 & * & 0 \\
0 & * & * \\
0 & * & *
\end{pmatrix}$

$\begin{pmatrix}
* & * & * \\
* & * & * \\
* & * & *
\end{pmatrix}$
&
$\begin{pmatrix}
0 & 0 & *\\
* & * & 0 \\
* & * & 0
\end{pmatrix}$

$\begin{pmatrix}
* & 0 & 0 \\
0 & * & * \\
0 & * & *
\end{pmatrix}$
\\ \hline

(\text{v},\text{ii})  
&
$\begin{pmatrix}
* & * & 0 \\
0 & 0 & 0 \\
0 & 0 & 0
\end{pmatrix}$

$\begin{pmatrix}
* & 0 & 0 \\
0 & * & * \\
0 & * & *
\end{pmatrix}$
&
$\begin{pmatrix}
0 & 0 & * \\
0 & 0 & * \\
0 & 0 & *
\end{pmatrix}$

$\begin{pmatrix}
* & * & * \\
* & * & * \\
* & * & *
\end{pmatrix}$
&
$\begin{pmatrix}
0 & 0 & 0\\
* & * & * \\
* & * & *
\end{pmatrix}$

$\begin{pmatrix}
* & 0 & 0 \\
0 & * & * \\
0 & * & *
\end{pmatrix}$
\\ \hline

(\text{v},\text{iii})  
&
$\begin{pmatrix}
* & * & 0 \\
0 & 0 & * \\
0 & 0 & *
\end{pmatrix}$

$\begin{pmatrix}
* & 0 & 0 \\
0 & * & * \\
0 & * & *
\end{pmatrix}$
&
$\begin{pmatrix}
0 & 0 & 0 \\
0 & 0 & * \\
0 & 0 & *
\end{pmatrix}$

$\begin{pmatrix}
* & * & * \\
* & * & * \\
* & * & *
\end{pmatrix}$
&
$\begin{pmatrix}
0 & 0 & *\\
* & * & 0 \\
* & * & 0
\end{pmatrix}$

$\begin{pmatrix}
* & 0 & 0 \\
0 & * & * \\
0 & * & *
\end{pmatrix}$
\\ \hline

(\text{v},\text{iv})  
&
$\begin{pmatrix}
* & 0 & 0 \\
0 & 0 & 0 \\
0 & 0 & 0
\end{pmatrix}$

$\begin{pmatrix}
* & 0 & 0 \\
0 & * & * \\
0 & * & *
\end{pmatrix}$
&
$\begin{pmatrix}
0 & * & * \\
0 & * & * \\
0 & * & *
\end{pmatrix}$

$\begin{pmatrix}
* & * & * \\
* & * & * \\
* & * & *
\end{pmatrix}$
&
$\begin{pmatrix}
0 & 0 & 0\\
* & * & * \\
* & * & *
\end{pmatrix}$

$\begin{pmatrix}
* & 0 & 0 \\
0 & * & * \\
0 & * & *
\end{pmatrix}$
\\ \hline

(\text{v},\text{v})  
&
$\begin{pmatrix}
* & 0 & 0 \\
0 & * & * \\
0 & * & *
\end{pmatrix}$

$\begin{pmatrix}
* & 0 & 0 \\
0 & * & * \\
0 & * & *
\end{pmatrix}$
&
$\begin{pmatrix}
0 & 0 & 0 \\
0 & * & * \\
0 & * & *
\end{pmatrix}$

$\begin{pmatrix}
* & * & * \\
* & * & * \\
* & * & *
\end{pmatrix}$
&
$\begin{pmatrix}
0 & * & *\\
* & 0 & 0 \\
* & 0 & 0
\end{pmatrix}$

$\begin{pmatrix}
* & 0 & 0 \\
0 & * & * \\
0 & * & *
\end{pmatrix}$
\\ \hline

(\text{v},\text{vi})  
&
$\begin{pmatrix}
0 & 0 & 0 \\
0 & 0 & * \\
0 & 0 & *
\end{pmatrix}$

$\begin{pmatrix}
* & 0 & 0 \\
0 & * & * \\
0 & * & *
\end{pmatrix}$
&
$\begin{pmatrix}
* & * & 0 \\
* & * & * \\
* & * & *
\end{pmatrix}$

$\begin{pmatrix}
* & * & * \\
* & * & * \\
* & * & *
\end{pmatrix}$
&
$\begin{pmatrix}
0 & 0 & *\\
* & * & 0 \\
* & * & 0
\end{pmatrix}$

$\begin{pmatrix}
* & 0 & 0 \\
0 & * & * \\
0 & * & *
\end{pmatrix}$
\\ \hline

(\text{v},\text{vii})  
&
$\begin{pmatrix}
0 & 0 & 0 \\
0 & * & * \\
0 & * & *
\end{pmatrix}$

$\begin{pmatrix}
* & 0 & 0 \\
0 & * & * \\
0 & * & *
\end{pmatrix}$
&
$\begin{pmatrix}
* & 0 & 0 \\
* & * & * \\
* & * & *
\end{pmatrix}$

$\begin{pmatrix}
* & * & * \\
* & * & * \\
* & * & *
\end{pmatrix}$
&
$\begin{pmatrix}
0 & * & *\\
* & 0 & 0 \\
* & 0 & 0
\end{pmatrix}$

$\begin{pmatrix}
* & 0 & 0 \\
0 & * & * \\
0 & * & *
\end{pmatrix}$
\\ \hline

(\text{vi},\text{i})  
&
$\begin{pmatrix}
0 & * & 0 \\
0 & * & 0 \\
0 & 0 & *
\end{pmatrix}$

$\begin{pmatrix}
* & * & 0 \\
* & * & 0 \\
0 & 0 & *
\end{pmatrix}$
&
$\begin{pmatrix}
* & 0 & * \\
* & 0 & * \\
0 & * & *
\end{pmatrix}$

$\begin{pmatrix}
* & * & * \\
* & * & * \\
* & * & *
\end{pmatrix}$
&
$\begin{pmatrix}
0 & * & * \\
0 & * & * \\
* & * & 0
\end{pmatrix}$

$\begin{pmatrix}
* & * & * \\
* & * & * \\
* & * & *
\end{pmatrix}$
\\ \hline

    \end{tabular}
    \caption{$Y^{ij}$ and ${C_{\mathrm{W}}}^{ij}$ matrices for $M=5$.}
    \label{tab:Y-C-M=5-3}
\end{table}

\begin{table}[H]
    \centering
    \begin{tabular}{|c|c|c|c|} \hline
        Flavor & Higgs\ $[g^0]$ & Higgs\ $[g^1]$ &Higgs\ $[g^2]$ \\ 
        (Left,Right) & $Y^{ij}$,\hspace{1cm} ${C_{\mathrm{W}}}^{ij}$ & $Y^{ij}$,\hspace{1cm} ${C_{\mathrm{W}}}^{ij}$ & $Y^{ij}$,\hspace{1cm} ${C_{\mathrm{W}}}^{ij}$\\\hline
(\text{vi},\text{ii})  
&
$\begin{pmatrix}
0 & 0 & * \\
0 & 0 & * \\
0 & 0 & 0
\end{pmatrix}$

$\begin{pmatrix}
* & * & 0 \\
* & * & 0 \\
0 & 0 & *
\end{pmatrix}$
&
$\begin{pmatrix}
* & * & 0 \\
* & * & 0 \\
0 & 0 & *
\end{pmatrix}$

$\begin{pmatrix}
* & * & * \\
* & * & * \\
* & * & *
\end{pmatrix}$
&
$\begin{pmatrix}
0 & 0 & * \\
0 & 0 & * \\
* & * & *
\end{pmatrix}$

$\begin{pmatrix}
* & * & * \\
* & * & * \\
* & * & *
\end{pmatrix}$
\\ \hline

(\text{vi},\text{iii})  
&
$\begin{pmatrix}
0 & 0 & 0 \\
0 & 0 & 0 \\
0 & 0 & *
\end{pmatrix}$

$\begin{pmatrix}
* & * & 0 \\
* & * & 0 \\
0 & 0 & *
\end{pmatrix}$
&
$\begin{pmatrix}
* & * & * \\
* & * & * \\
0 & 0 & *
\end{pmatrix}$

$\begin{pmatrix}
* & * & * \\
* & * & * \\
* & * & *
\end{pmatrix}$
&
$\begin{pmatrix}
0 & 0 & * \\
0 & 0 & * \\
* & * & 0
\end{pmatrix}$

$\begin{pmatrix}
* & * & * \\
* & * & * \\
* & * & *
\end{pmatrix}$
\\ \hline

(\text{vi},\text{iv})  
&
$\begin{pmatrix}
0 & * & * \\
0 & * & * \\
0 & 0 & 0
\end{pmatrix}$

$\begin{pmatrix}
* & * & 0 \\
* & * & 0 \\
0 & 0 & *
\end{pmatrix}$
&
$\begin{pmatrix}
* & 0 & 0 \\
* & 0 & 0 \\
0 & * & *
\end{pmatrix}$

$\begin{pmatrix}
* & * & * \\
* & * & * \\
* & * & *
\end{pmatrix}$
&
$\begin{pmatrix}
0 & * & * \\
0 & * & * \\
* & * & *
\end{pmatrix}$

$\begin{pmatrix}
* & * & * \\
* & * & * \\
* & * & *
\end{pmatrix}$
\\ \hline

(\text{vi},\text{v})  
&
$\begin{pmatrix}
0 & 0 & 0 \\
0 & 0 & 0 \\
0 & * & *
\end{pmatrix}$

$\begin{pmatrix}
* & * & 0 \\
* & * & 0 \\
0 & 0 & *
\end{pmatrix}$
&
$\begin{pmatrix}
* & * & * \\
* & * & * \\
0 & * & *
\end{pmatrix}$

$\begin{pmatrix}
* & * & * \\
* & * & * \\
* & * & *
\end{pmatrix}$
&
$\begin{pmatrix}
0 & * & * \\
0 & * & * \\
* & 0 & 0
\end{pmatrix}$

$\begin{pmatrix}
* & * & * \\
* & * & * \\
* & * & *
\end{pmatrix}$
\\ \hline

(\text{vi},\text{vi})  
&
$\begin{pmatrix}
* & * & 0 \\
* & * & 0 \\
0 & 0 & *
\end{pmatrix}$

$\begin{pmatrix}
* & * & 0 \\
* & * & 0 \\
0 & 0 & *
\end{pmatrix}$
&
$\begin{pmatrix}
0 & 0 & * \\
0 & 0 & * \\
* & * & *
\end{pmatrix}$

$\begin{pmatrix}
* & * & * \\
* & * & * \\
* & * & *
\end{pmatrix}$
&
$\begin{pmatrix}
* & * & * \\
* & * & * \\
* & * & 0
\end{pmatrix}$

$\begin{pmatrix}
* & * & * \\
* & * & * \\
* & * & *
\end{pmatrix}$
\\ \hline

(\text{vi},\text{vii})  
&
$\begin{pmatrix}
* & 0 & 0 \\
* & 0 & 0 \\
0 & * & *
\end{pmatrix}$

$\begin{pmatrix}
* & * & 0 \\
* & * & 0 \\
0 & 0 & *
\end{pmatrix}$
&
$\begin{pmatrix}
0 & * & * \\
0 & * & * \\
* & * & *
\end{pmatrix}$

$\begin{pmatrix}
* & * & * \\
* & * & * \\
* & * & *
\end{pmatrix}$
&
$\begin{pmatrix}
* & * & * \\
* & * & * \\
* & 0 & 0
\end{pmatrix}$

$\begin{pmatrix}
* & * & * \\
* & * & * \\
* & * & *
\end{pmatrix}$
\\ \hline

(\text{vii},\text{i})  
&
$\begin{pmatrix}
0 & * & 0 \\
0 & 0 & * \\
0 & 0 & *
\end{pmatrix}$

$\begin{pmatrix}
* & 0 & 0 \\
0 & * & * \\
0 & * & *
\end{pmatrix}$
&
$\begin{pmatrix}
* & 0 & * \\
0 & * & * \\
0 & * & *
\end{pmatrix}$

$\begin{pmatrix}
* & * & * \\
* & * & * \\
* & * & *
\end{pmatrix}$
&
$\begin{pmatrix}
0 & * & * \\
* & * & 0 \\
* & * & 0
\end{pmatrix}$

$\begin{pmatrix}
* & * & * \\
* & * & * \\
* & * & *
\end{pmatrix}$
\\ \hline

(\text{vii},\text{ii})  
&
$\begin{pmatrix}
0 & 0 & * \\
0 & 0 & 0 \\
0 & 0 & 0
\end{pmatrix}$

$\begin{pmatrix}
* & 0 & 0 \\
0 & * & * \\
0 & * & *
\end{pmatrix}$
&
$\begin{pmatrix}
* & * & 0 \\
0 & 0 & * \\
0 & 0 & *
\end{pmatrix}$

$\begin{pmatrix}
* & * & * \\
* & * & * \\
* & * & *
\end{pmatrix}$
&
$\begin{pmatrix}
0 & 0 & * \\
* & * & * \\
* & * & *
\end{pmatrix}$

$\begin{pmatrix}
* & * & * \\
* & * & * \\
* & * & *
\end{pmatrix}$
\\ \hline

(\text{vii},\text{iii})  
&
$\begin{pmatrix}
0 & 0 & 0 \\
0 & 0 & * \\
0 & 0 & *
\end{pmatrix}$

$\begin{pmatrix}
* & 0 & 0 \\
0 & * & * \\
0 & * & *
\end{pmatrix}$
&
$\begin{pmatrix}
* & * & * \\
0 & 0 & * \\
0 & 0 & *
\end{pmatrix}$

$\begin{pmatrix}
* & * & * \\
* & * & * \\
* & * & *
\end{pmatrix}$
&
$\begin{pmatrix}
0 & 0 & * \\
* & * & 0 \\
* & * & 0
\end{pmatrix}$

$\begin{pmatrix}
* & * & * \\
* & * & * \\
* & * & *
\end{pmatrix}$
\\ \hline

(\text{vii},\text{iv})  
&
$\begin{pmatrix}
0 & * & * \\
0 & 0 & 0 \\
0 & 0 & 0
\end{pmatrix}$

$\begin{pmatrix}
* & 0 & 0 \\
0 & * & * \\
0 & * & *
\end{pmatrix}$
&
$\begin{pmatrix}
* & 0 & 0 \\
0 & * & * \\
0 & * & *
\end{pmatrix}$

$\begin{pmatrix}
* & * & * \\
* & * & * \\
* & * & *
\end{pmatrix}$
&
$\begin{pmatrix}
0 & * & * \\
* & * & * \\
* & * & *
\end{pmatrix}$

$\begin{pmatrix}
* & * & * \\
* & * & * \\
* & * & *
\end{pmatrix}$
\\ \hline

(\text{vii},\text{v})  
&
$\begin{pmatrix}
0 & 0 & 0 \\
0 & * & * \\
0 & * & *
\end{pmatrix}$

$\begin{pmatrix}
* & 0 & 0 \\
0 & * & * \\
0 & * & *
\end{pmatrix}$
&
$\begin{pmatrix}
* & * & * \\
0 & * & * \\
0 & * & *
\end{pmatrix}$

$\begin{pmatrix}
* & * & * \\
* & * & * \\
* & * & *
\end{pmatrix}$
&
$\begin{pmatrix}
0 & * & * \\
* & 0 & 0 \\
* & 0 & 0
\end{pmatrix}$

$\begin{pmatrix}
* & * & * \\
* & * & * \\
* & * & *
\end{pmatrix}$
\\ \hline

(\text{vii},\text{vi})  
&
$\begin{pmatrix}
* & * & 0 \\
0 & 0 & * \\
0 & 0 & *
\end{pmatrix}$

$\begin{pmatrix}
* & 0 & 0 \\
0 & * & * \\
0 & * & *
\end{pmatrix}$
&
$\begin{pmatrix}
0 & 0 & * \\
* & * & * \\
* & * & *
\end{pmatrix}$

$\begin{pmatrix}
* & * & * \\
* & * & * \\
* & * & *
\end{pmatrix}$
&
$\begin{pmatrix}
* & * & * \\
* & * & 0 \\
* & * & 0
\end{pmatrix}$

$\begin{pmatrix}
* & * & * \\
* & * & * \\
* & * & *
\end{pmatrix}$
\\ \hline

    \end{tabular}
    \caption{$Y^{ij}$ and ${C_{\mathrm{W}}}^{ij}$ matrices for $M=5$.}
    \label{tab:Y-C-M=5-4}
\end{table}

\begin{table}[H]
    \centering
    \begin{tabular}{|c|c|c|c|} \hline
        Flavor & Higgs\ $[g^0]$ & Higgs\ $[g^1]$ &Higgs\ $[g^2]$ \\ 
        (Left,Right) & $Y^{ij}$,\hspace{1cm} ${C_{\mathrm{W}}}^{ij}$ & $Y^{ij}$,\hspace{1cm} ${C_{\mathrm{W}}}^{ij}$ & $Y^{ij}$,\hspace{1cm} ${C_{\mathrm{W}}}^{ij}$\\\hline

(\text{vii},\text{vii})  
&
$\begin{pmatrix}
* & 0 & 0 \\
0 & * & * \\
0 & * & *
\end{pmatrix}$

$\begin{pmatrix}
* & 0 & 0 \\
0 & * & * \\
0 & * & *
\end{pmatrix}$
&
$\begin{pmatrix}
0 & * & * \\
* & * & * \\
* & * & *
\end{pmatrix}$

$\begin{pmatrix}
* & * & * \\
* & * & * \\
* & * & *
\end{pmatrix}$
&
$\begin{pmatrix}
* & * & * \\
* & 0 & 0 \\
* & 0 & 0
\end{pmatrix}$

$\begin{pmatrix}
* & * & * \\
* & * & * \\
* & * & *
\end{pmatrix}$
\\ \hline

    \end{tabular}
    \caption{$Y^{ij}$ and ${C_{\mathrm{W}}}^{ij}$ matrices for $M=5$.}
    \label{tab:Y-C-M=5-5}
\end{table}

\subsubsection{Classification}

Here, we classify the results in sections  \ref{sec:Y-C-M=3}, \ref{sec:Y-C-M=4}, and \ref{sec:Y-C-M=5}.
The possible ${C_{\mathrm{W}}}^{ij}$ matrices are classified to the following four patterns:
\begin{align}
 C_{\mathrm{W}}^{ij(\mathrm{1})} = \begin{pmatrix}
  * & 0 & 0 \\
  0 & * & 0 \\
  0 & 0 & * \\
\end{pmatrix}, \qquad    C_{\mathrm{W}}^{ij(\mathrm{2})} = 
\begin{pmatrix}
  * & * & 0 \\
  * & * & 0 \\
  0 & 0 & * \\
\end{pmatrix}, \qquad 
C_{\mathrm{W}}^{ij(\mathrm{3})} = \begin{pmatrix}
  * & 0 & * \\
  0 & * & * \\
  * & * & * \\
\end{pmatrix}, \qquad  
C_{\mathrm{W}}^{ij(\mathrm{4})} = \begin{pmatrix}
  * & * & * \\
  * & * & * \\
  * & * & * \\
\end{pmatrix},
\end{align}
as well as their permutations of rows and columns.
Several combinations between $Y^{ij}$ and $C_{\mathrm{W}}^{ij(k)}$ are realized for each $C_{\mathrm{W}}^{ij(k)}$.
Possible $Y^{ij}$ textures for $C_{\mathrm{W}}^{ij(k)}$ with $k=1,2,3,4$ are shown in Tables \ref{tab:YforC1}, \ref{tab:YforC2}, \ref{tab:YforC3}, and \ref{tab:YforC4}, where Yukawa textures with ${\rm det}~Y^{ij}=0$ are omitted, because they lead to at least one massless charged lepton.
Indeed, there are many textures with ${\rm det}~Y^{ij}=0$.
Also in these tables, possible permutations of rows and columns in $Y^{ij}$ matrix  are omitted.

For $M=3$, the diagonal neutrino mass matrix $C_{\mathrm{W}}^{ij(1)}$ can not be realized.
For $M=4$ and $C_{\mathrm{W}}^{ij(1)}$, two or three mixing angles are vanishing, and these cases are not realistic.
Similarly, the first two textures for $M=5$ can not lead to all of three non-vanishing mixing angles.
The other two textures $Y^{ij}$ with four zeros and three zeros would be interesting.
Note that these two textures $Y^{ij}$ can not be realized by conventional group-like $\mathbb{Z}_n$ symmetries, if they are not spontaneously broken. 
We will study them in the next section.

$C_{\mathrm{W}}^{ij(2)}$ can be realized for $M=3,4,5$.
Similarly to $C_{\mathrm{W}}^{ij(1)}$, for $M=4$ and $C_{\mathrm{W}}^{ij(2)}$, two or three mixing angles are vanishing.
Also, the first texture for $M=3$ and 5 can not lead to all of three non-vanishing mixing angles.
Thus, these cases are not realistic.
Among the others, the Yukawa texture $Y^{ij}$ with three zeros includes most zeros and it is obtained when $M=5$.
We will study this case in the next section so as to show that this case is realistic.
Note that this texture $Y^{ij}$ can not be realized by conventional group-like $\mathbb{Z}_n$ symmetries.
The Yukawa textures $Y^{ij}$ with two zeros and one zero include more free parameters.
One can tune those parameters to realize experimental data.

$C_{\mathrm{W}}^{ij(3)}$ can be realized only for $M=5$.
$C_{\mathrm{W}}^{ij(4)}$ can be realized for $M=3,4,5$.
These neutrino mass textures include more free parameters.
For example, $C_{\mathrm{W}}^{ij(3)}$ has five free parameters, which are sufficient to explain the difference of neutrino masses squared and three mixing angles even if $Y^{ij}$ is diagonal.
Indeed, the neutrino mass matrix $C_{\mathrm{W}}^{ij(3)}$ with vanishing (1,1) entry corresponds to the so-called $A_1$ texture in Ref.~\cite{Frampton:2002yf}, and it is realistic.

\begin{table}[H]
    \centering
\begin{tabular}{|l|cccc|}
\hline
$M$  & \multicolumn{4}{c|}{Yukawa textures $Y^{ij}$for $C_{\mathrm{W}}^{ij(\mathrm{1})}$ } \\
\hline
\ $4$& 
$\begin{pmatrix}
  * & 0 & 0 \\
  0 & * & 0 \\
  0 & 0 & * \\
\end{pmatrix}$&
$\begin{pmatrix}
  * & * & 0 \\
  * & * & 0 \\
  0 & 0 & * \\
\end{pmatrix}$ &  & \\
\hline
\ $5$ & 
$\begin{pmatrix}
  * & 0 & 0 \\
  0 & * & 0 \\
  0 & 0 & * \\
\end{pmatrix}$&
$\begin{pmatrix}
  * & * & 0 \\
  * & * & 0 \\
  0 & 0 & * \\
\end{pmatrix}$&
$\begin{pmatrix}
  0 & 0 & * \\
  0 & * & * \\
  * & * & 0 \\
\end{pmatrix}$&
$\begin{pmatrix}
  * & 0 & * \\
  * & * & * \\
  0 & * & 0 \\
\end{pmatrix}$ \\
\hline
\end{tabular}
   \caption{Possible $Y^{ij}$ for $C_{\mathrm{W}}^{ij(\mathrm{1})}$ with $M=4,5$.}
    \label{tab:YforC1}
\end{table}

\begin{table}[H]
    \centering
\begin{tabular}{|l|ccccc|}
\hline
$M$  & \multicolumn{5}{c|}{Yukawa textures $Y^{ij}$for $C_{\mathrm{W}}^{ij(\mathrm{2})}$} \\
\hline
\ $3$ & 
$\begin{pmatrix}
  * & * & 0 \\
  * & * & 0 \\
  0 & 0 & * \\
\end{pmatrix}$& 
$\begin{pmatrix}
  * & * & * \\
  * & * & * \\
  * & 0 & 0 \\
\end{pmatrix}$&
$\begin{pmatrix}
  * & 0 & * \\
  * & 0 & * \\
  * & * & * \\
\end{pmatrix}$&
$\begin{pmatrix}
  * & * & * \\
  * & * & * \\
  * & * & 0 \\
\end{pmatrix}$&\\
\hline
\ $4$ & 
$\begin{pmatrix}
  * & 0 & 0 \\
  0 & * & 0 \\
  0 & 0 & * \\
\end{pmatrix}$&
$\begin{pmatrix}
  * & * & 0 \\
  * & * & 0 \\
  0 & 0 & * \\
\end{pmatrix}$ & & & \\
\hline
\ $5$ & 
$\begin{pmatrix}
  * & * & 0 \\
  * & * & 0 \\
  0 & 0 & * \\
\end{pmatrix}$&
$\begin{pmatrix}
  * & * & 0 \\
  * & * & 0 \\
  0 & * & * \\
\end{pmatrix}$&
$\begin{pmatrix}
  * & 0 & * \\
  * & 0 & * \\
  * & * & * \\
\end{pmatrix}$&
$\begin{pmatrix}
  * & * & * \\
  * & * & * \\
  * & 0 & 0 \\
\end{pmatrix}$&
$\begin{pmatrix}
  * & * & * \\
  * & * & * \\
  * & * & 0 \\
\end{pmatrix}$\\
\hline
\end{tabular}
  \caption{Possible $Y^{ij}$ for $C_{\mathrm{W}}^{ij(\mathrm{2})}$ with $M=3,4,5$.}
    \label{tab:YforC2}
\end{table}

\begin{table}[H]
    \centering
\begin{tabular}{|l|ccccc|}
\hline
$M$  & \multicolumn{5}{c|}{Yukawa textures $Y^{ij}$for $C_{\mathrm{W}}^{ij(\mathrm{3})}$ } \\
\hline
\ $5$ & 
$\begin{pmatrix}
  * & 0 & 0 \\
  0 & * & 0 \\
  0 & 0 & * \\
\end{pmatrix}$&
$\begin{pmatrix}
  * & * & 0 \\
  * & * & 0 \\
  0 & 0 & * \\
\end{pmatrix}$&
$\begin{pmatrix}
  0 & * & * \\
  * & 0 & 0 \\
  0 & * & * \\
\end{pmatrix}$&
$\begin{pmatrix}
  0 & 0 & * \\
  0 & * & * \\
  * & * & 0 \\
\end{pmatrix}$&
$\begin{pmatrix}
  0 & 0 & * \\
  * & * & 0 \\
  0 & * & * \\
\end{pmatrix}$ \\
&$\begin{pmatrix}
  0 & * & * \\
  0 & 0 & * \\
  * & * & 0 \\
\end{pmatrix}$&
$\begin{pmatrix}
  * & 0 & * \\
  * & * & * \\
  0 & * & 0 \\
\end{pmatrix}$&
$\begin{pmatrix}
  * & 0 & 0 \\
  * & * & * \\
  0 & * & * \\
\end{pmatrix}$&
$\begin{pmatrix}
  * & 0 & * \\
  0 & * & 0 \\
  * & * & * \\
\end{pmatrix}$ &\\
\hline
\end{tabular}
  \caption{Possible $Y^{ij}$ for $C_{\mathrm{W}}^{ij(\mathrm{3})}$ with $M=5$.}
    \label{tab:YforC3}
\end{table}

\begin{table}[H]
    \centering
\begin{tabular}{|l|ccccc|}
\hline
$M$  & \multicolumn{5}{c|}{Yukawa textures $Y^{ij}$ for $C_{\mathrm{W}}^{ij(\mathrm{4})}$} \\
\hline
\ $3$ & 
$\begin{pmatrix}
  * & * & 0 \\
  * & * & 0 \\
  0 & 0 & * \\
\end{pmatrix}$&
$\begin{pmatrix}
  * & * & * \\
  * & * & * \\
  * & 0 & 0 \\
\end{pmatrix}$&
$\begin{pmatrix}
  * & * & * \\
  * & 0 & * \\
  * & 0 & * \\
\end{pmatrix}$&
$\begin{pmatrix}
  * & * & * \\
  * & * & * \\
  * & * & 0 \\
\end{pmatrix}$ &\\
\hline
\ $4$ & 
$\begin{pmatrix}
  * & * & 0 \\
  * & * & 0 \\
  0 & 0 & * \\
\end{pmatrix}$ & & & & \\
\hline
\ $5$ & 
$\begin{pmatrix}
  * & * & 0 \\
  * & * & 0 \\
  0 & 0 & * \\
\end{pmatrix}$&
$\begin{pmatrix}
  * & * & 0 \\
  0 & * & * \\
  * & * & 0 \\
\end{pmatrix}$&
$\begin{pmatrix}
  * & 0 & * \\
  * & * & * \\
  * & 0 & * \\
\end{pmatrix}$&
$\begin{pmatrix}
  * & * & * \\
  * & * & * \\
  * & 0 & 0 \\
\end{pmatrix}$&
$\begin{pmatrix}
  * & * & * \\
  * & * & * \\
  * & * & 0 \\
\end{pmatrix}$ \\
\hline
\end{tabular}
  \caption{Possible $Y^{ij}$ for $C_{\mathrm{W}}^{ij(\mathrm{4})}$ with $M=3,4,5$.}
    \label{tab:YforC4}
\end{table}

\section{Phenomenological aspects}
\label{sec:pheno}

In the previous sections, we have derived mass textures in the lepton sector from our selection rules.
Here, we study their phenomenological aspects.

For example, 
in Ref.~\cite{Frampton:2002yf}, the neutrino mass textures   were studied, i.e., textures $A_{1,2}$, $B_{1,2,3,4}$, $C$, 
which include texture zeros in the diagonal entries in the basis that the charged lepton mass matrix is diagonal.
Our selection rules can not derive those textures, but diagonal entries in the neutrino mass matrix are always allowed.
Among neutrino mass matrices with non-vanishing diagonal entries, the texture with {the} most zeros {is} the diagonal one,
\begin{align}
\label{eq:C-diag}
    {C_{\mathrm{W}}}^{ij}=
    \begin{pmatrix}
        * & 0 & 0\\
        0 & * & 0\\
        0 & 0 & *
    \end{pmatrix}.
\end{align}
It can lead to neutrino masses by choosing free parameters properly.

When ${C_{\mathrm{W}}}^{ij}$ is diagonal, the Yukawa matrix $Y^{ij}$ in the charged lepton sector must realize experimental values of charged lepton masses \cite{ParticleDataGroup:2024cfk}: 
\begin{align}
m_e=0.511 ~{\rm MeV},\qquad m_\mu = 106 ~{\rm MeV}, \qquad m_\tau=1777~{\rm MeV},
\end{align}
and the PMNS matrix,
\begin{align}
U=
\begin{pmatrix}
 1 & 0 & 0\\
0 & c_{23} & s_{23} \\
0 & -s_{23} & c_{23}
\end{pmatrix}
\begin{pmatrix}
 c_{13} & 0 & s_{13}e^{-i\delta_{\mathrm{CP}}}\\
0 & 1 & 0 \\
-s_{13}e^{i\delta_{\mathrm{CP}}} & 0 & c_{13}
\end{pmatrix}
\begin{pmatrix}
 c_{12} & s_{12} & 0\\
 -s_{12} & c_{12} & 0 \\
0 & 0 & 1
\end{pmatrix},
\end{align}
where $c_{ij}=\cos \theta_{ij}$ and $s_{ij}=\sin \theta_{ij}$.
In the normal hierarchy (NH), the experimental values of the PMNS matrix are \cite{Esteban:2024eli}
\begin{align}
\begin{array}{ll}
 \sin^2\theta_{12}=0.308^{+0.012}_{-0.011}\,, 
 &\theta_{12}/ ^{\circ}=33.68^{+0.73}_{-0.70}\,,  \\
 \sin^2\theta_{23}=0.470^{+0.017}_{-0.013}\,, &\theta_{23}/^{\circ}=43.3^{+1.0}_{-0.8}\,, \\
 \sin^2\theta_{13}=0.02215^{+0.00056}_{-0.00058}\,, &\theta_{13}/^{\circ}=8.56^{+0.11}_{-0.11}\,,\\
 {\delta_{\mathrm{CP}}}/^{\circ}=212^{+26}_{-41}\,.
\end{array}
\end{align}
That is, the Yukawa matrix must satisfy 
\begin{align}
    YY^\dagger v^2=U^{-1}\begin{pmatrix}
m_e^2 & 0 & 0 \\
0 & m_\mu^2 & 0 \\
0 & 0& m_\tau^2
\end{pmatrix}U\,,
\end{align}
where $v$ is the vacuum expectation value of $H$ in the Standard Model and it is replaced by the vacuum expectation value of $H_d$ in supersymmetric standard models and two Higgs doublet models.

Table~\ref{tab:Y-C-M=5-1} shows that the following Yukawa textures:
\begin{align}
    Y_{(a)}=
    \begin{pmatrix}
        0 & * & 0 \\
        * & 0 & * \\
        0 & * & *
    \end{pmatrix}, \qquad 
    Y_{(b)}=
    \begin{pmatrix}
        0 & 0 & * \\
        * & * & * \\
        * & * & 0
    \end{pmatrix},  
\end{align}
including their permutations as well as matrices with vanishing determinants  
can be obtained in combination with the diagonal ${C_{\mathrm{W}}}^{ij}$ of Eq.~(\ref{eq:C-diag}).
The first one $Y_{(a)}$ is the nearest neighbor interaction texture. 
This matrix is real one after removing the unphysical phases by rephasing fermion fields since ${C_{\mathrm{W}}}^{ij}$ is diagonal. Therefore, the CP symmetry is conserved in this texture.
One can also easily see $Y_{(b)}$ being the conserved CP
by calculating the Jarlskog invariant directly \cite{Jarlskog:1985ht}.

Furthermore, when we combine two selection rules, i.e., Deligne tensor product of two Fibonacci categories, we can derive the following texture:
\begin{align}
\label{eq:texture_quark}
    Y_{(c)}=
    \begin{pmatrix}
        0 & 0 & * \\
        0 & * & * \\
        * & * & *
    \end{pmatrix}, 
\end{align}
and the diagonal ${C_{\mathrm{W}}}^{ij}$.
For example, we assign the classes of the first symmetry with $M=3$ to $L_i$ and $e_i$ as 
\begin{align}
    L_i:([g^0],[g^1],[g^1]),\qquad e_i:([g^0],[g^0],[g^1]),
\end{align}
where $H_u$ and $H_d$ have $[g^0]$ and $[g^1]$.
On top of that, let us assign the classes of the second symmetry with $M=3$ as 
\begin{align}
    L_i:([g^1],[g^0],[g^1]),\qquad e_i:([g^0],[g^1],[g^1]),
\end{align}
where $H_u$ and $H_d$ have $[g^0]$ and $[g^1]$.
These combinations lead to the above $Y_{(c)}$ and the diagonal ${C_{\mathrm{W}}}^{ij}$.
Then, we take the parameters as 
\begin{align}
    Y_{(c)}=Y_\tau
    \begin{pmatrix}
     0  & 0 & 0.64\\
     0  & 2.7 \times 10^{-2} e^{ 142^\circ\, i}& {0.73}\\
    2.7\times 10^{-3}   & {0.10}  & 1
    \end{pmatrix},
\end{align}
so as to obtain the experimental values of the charged lepton masses and the PMNS matrix in the case of NH of neutrino masses.
For the case of the inverted hierarchy (IH) of neutrino masses, this texture is completely consistent with observed one \cite{Esteban:2024eli}.

Next, let us discuss another texture of ${C_{\mathrm{W}}}^{ij}$.
The following texture of ${C_{\mathrm{W}}}^{ij}$:
\begin{align}
\label{eq:C-(1,2)}
    {C_{\mathrm{W}}}^{ij}=
    \begin{pmatrix}
        * & * &0 \\
        * & * &0 \\
        0 & 0 & *
    \end{pmatrix}
\end{align}
includes the next most zeros after the diagonal one.
It can realize the neutrino masses and the (1,2) mixing angle by choosing proper values of parameters.

For example, we consider the flavor (vii,i) in Table~\ref{tab:Y-C-M=5-4} with $M=5$.
Also, we change the ordering of $L_i$ from $[g^1][g^2][g^2]$ to $[g^2][g^2][g^1]$.
That is, we exchange the first and third generations of $L_i$.
Then, we realize the above ${C_{\mathrm{W}}}^{ij}$ 
in Eq.~(\ref{eq:C-(1,2)}) when the Higgs field $H_u$ corresponds to $[g^0]$.
On top of that, when the Higgs field $H_d$ corresponds to $[g^1]$, we obtain the following Yukawa texture:
\begin{align}
    Y_{(d)}=
    \begin{pmatrix}
    0& * & * \\
    0 & * & * \\
    * & 0 & *
    \end{pmatrix},
\end{align}
in the charged lepton sector.
Also, the permutations of $Y_{(d)}$ can be derived.
When we choose the parameters in ${C_{\mathrm{W}}}^{ij}$ of Eq.~(\ref{eq:C-(1,2)}) such that they lead to the correct value of (1,2) mixing angle, we tune the parameters of $Y_{(e)}$ such that they satisfy 
\begin{align}
    Y_{(d)}Y^\dagger_{(d)} v^2_d=U'^{-1}\begin{pmatrix}
m_e^2 & 0 & 0 \\
0 & m_\mu^2 & 0 \\
0 & 0& m_\tau^2
\end{pmatrix}U',
\end{align}
where 
\begin{align}
U'=
\begin{pmatrix}
 1 & 0 & 0\\
0 & c_{23} & s_{23} \\
0 & -s_{23} & c_{23}
\end{pmatrix}
\begin{pmatrix}
 c_{13} & 0 & s_{13}e^{-i\delta_{CP}}\\
0 & 1 & 0 \\
-s_{13}e^{i\delta_{CP}} & 0 & c_{13}
\end{pmatrix}.
\end{align}
We find a parameter set:
\begin{align}
    Y_{(d)}=Y_\tau
    \begin{pmatrix}
     0  & 8.8 \times 10^{-6}  & 0.15 \\
     0  & 0.11 e^{{110^\circ}\,i}& 0.95 \\
    2.7\times 10^{-3}   & 0 & 1
    \end{pmatrix}
\end{align}
satisfies the above condition to realize the charged lepton masses and the PMNS matrix for NH except the (1,2) mixing angle, which is realized by ${C_{\mathrm{W}}}^{ij}$ in Eq.~(\ref{eq:C-(1,2)}). 
This texture is also consistent with the observed mixing angles and the CP phase for the IH case.

\section{Conclusions}
\label{sec:con}

We have studied the lepton mass textures, which are derived by $\mathbb{Z}_2$ gauging of $\mathbb{Z}_M$ symmetries.
We have obtained various textures for the Yukawa couplings in the charged lepton sector, but the patterns of neutrino mass matrices are limited.
The reason why neutrino mass textures are limited is that diagonal entries are always allowed by our selection rules.
At any rate, all the obtained textures can not obtained by group-theoretical symmetries, and certain textures can lead to realistic results.

The textures in the quark sector have been studied in Refs.~\cite{Kobayashi:2024cvp,Kobayashi:2025znw}.
It is very important to combine the analyses in the quark and lepton sectors. For instance, the texture \eqref{eq:texture_quark} is also available to explain the masses and mixings in the quark sector, which can address the strong CP problem without axion \cite{Kobayashi:2025znw}. 
By such analysis, we could discuss flavor physics appearing as higher-dimensional operators in the standard model effective field theory.
It is also interesting to study the grand unified theory of quarks and leptons.
We would study them elsewhere.

Here we have studied $\mathbb{Z}_2$ gauging of $\mathbb{Z}_M$ symmetries, leading to new coupling selection rules including the Fibonacci and Ising fusion rules in the lepton sector.
It is interesting to apply other non-invertible selection rules to derive specific textures.\footnote{For example, Calabi-Yau compactifications lead to non-invertible fusion rules \cite{Dong:2025pah}.}
We may have other coupling selection rules, which forbid some of diagonal entries of the neutrino mass matrix, although our selection rules always allow them.

\acknowledgments

This work was supported by JSPS KAKENHI Grant Numbers JP23K03375 (T.K.) and JP25H01539 (H.O.).

\bibliography{references}{}
\bibliographystyle{JHEP}

\end{document}